\documentclass[smallextended,numbook]{svjour3}

\journalname{}

\usepackage{ulem}

\usepackage{graphicx,float,amsmath,amssymb,mathrsfs}

\usepackage[english,francais]{babel}

\usepackage{url,hyperref}  
\usepackage[usenames,dvipsnames]{color}

\renewcommand{\em}{\it}

\definecolor{M_Beige}         {rgb}{0.96 , 0.96 , 0.86}

\definecolor{M_Brown}         {rgb}{0.65 , 0.16 , 0.16}

\definecolor{M_Gold}          {rgb}{1.00 , 0.84 , 0.00}

\definecolor{M_LemonChiffon}  {rgb}{1.00 , 0.98 , 0.80}

\definecolor{M_Orange}        {rgb}{1.00 , 0.60 , 0.00}

\definecolor{M_Pink}          {rgb}{0.80 , 0.55 , 0.60}

\definecolor{M_Violet}          {rgb}{0.83 , 0.21 , 0.93}

\definecolor{M_Green}          {rgb}{0.2 , 0.6 , 0.2}

\definecolor{M_Gray}          {rgb}{0.7 , 0.7 , 0.7}

\definecolor{M_BluPal}          {rgb}{0.7 , 0.7 , 0.9}

\renewcommand{\geq}{\geqslant}

\newcommand{\ket}[1]{|\kern.3ex#1\kern.3ex\rangle}
\newcommand{\bra}[1]{\langle\kern.3ex #1 \kern.3ex|}
\newcommand{\scalar}[2]{\langle\kern.3ex{#1}\kern.3ex|\kern.3ex{#2}\kern.3ex\rangle}
\newcommand{\mean}[1]{\left\langle #1\right\rangle}
\newcommand{\smean}[1]{\langle #1\rangle}
\def\e{\mathrm{e}} % e de l'exponentielle

\newcommand{\tr}[1]{ \mathop{\mathrm{tr}}\nolimits\left\{ #1 \right\} }  % Trace
\newcommand{\heaviside}{\mathop{\theta_\mathrm{H}}\nolimits}  

\def\I{{\rm i}}                  % le i mathematique 
\def\dd{{\rm d}}                  % la differenciation
\def\D{{\mathcal{D}}}                 % D integrale fonctionnelle

\newcommand{\deriv}[2]{\frac{\mathrm{d}#1}{\mathrm{d}#2}}
\newcommand{\derivp}[2]{\frac{\partial #1}{\partial #2}}

\def\Xint#1{\mathchoice
{\XXint\displaystyle\textstyle{#1}}%
{\XXint\textstyle\scriptstyle{#1}}%
{\XXint\scriptstyle\scriptscriptstyle{#1}}%
{\XXint\scriptscriptstyle\scriptscriptstyle{#1}}%
\!\int}
\def\XXint#1#2#3{{\setbox0=\hbox{$#1{#2#3}{\int}$}
\vcenter{\hbox{$#2#3$}}\kern-.5\wd0}}

\def\dashint{\Xint-}

\newcommand{\abs}[1]{\left| #1 \right|}

\newcommand\antiddots{\mathinner{\mkern2mu\raise1pt\hbox{.}\mkern2mu
    \newline \raise4pt\hbox{.}\mkern2mu\raise7pt\hbox{.}\mkern1mu}}

\def\P{P_{N,\kappa}}
\def\Sm{\mathcal{S}}
\def\WSm{\mathcal{Q}}

\def\Ht{\tau_\mathrm{H}}

\def\Entropy{S}

\def\Var{\mathrm{Var}}
\def\Cov{\mathrm{Cov}}

\def\Nt{p}

\def\O{\mathcal{O}}

\def\G{G_{N,\kappa}}
\def\Z{Z_{N,p}}
\def\f{F}

\def\t{\tilde{\beta}}
\def\ts{\t_\mathrm{mic}}
\def\zt{z_\mathrm{can}}
\def\zs{z_\mathrm{mic}}

\def\rhoMP{\rho_0^\star}

\def\Sc{\bar{s}}

\def\a{l}
\def\m{n}

\def\ConstN{\delta_{\Nt,\sum_i \m_i}}

\def\cm{c_-}
\def\cp{c_+}
\def\cpt{\tilde{c}_+}

\begin{document}

\selectlanguage{english}

%%%%%%%%%%%%%%%%%%%%%%%%%%%%%%%%%%%%%%%%%%%%%%%%%%%%%%%%%%%%%%%%%%%%%%%%%%%%%
\renewcommand{\labelitemi}{$\bullet$}
\renewcommand{\labelitemii}{$\star$}
%%%%%%%%%%%%%%%%%%%%%%%%%%%%%%%%%%%%%%%%%%%%%%%%%%%%%%%%%%%%%%%%%%%%%%%%%%%%%

\title{Truncated linear statistics associated with the eigenvalues of random matrices II. Partial sums over proper time delays for chaotic quantum dots}

\author{Aur\'elien Grabsch \and Satya N. Majumdar \and Christophe Texier}

\institute{
Aur\'elien Grabsch \and Satya N. Majumdar \and Christophe Texier
\at
LPTMS, CNRS, Univ. Paris-Sud, Universit\'e Paris-Saclay, 91405 Orsay, France.
}

%\date{\today}
%\date{December 16, 2016}
\date{March 24, 2017}

\maketitle

\begin{abstract}
Invariant ensembles of random matrices are characterized by the distribution of their eigenvalues $\{\lambda_1,\cdots,\lambda_N\}$.
We study the distribution of truncated linear statistics of the form $\tilde{L}=\sum_{i=1}^p f(\lambda_i)$ with $p<N$. 
This problem has been considered by us in a previous paper when the $p$ eigenvalues are further constrained to be the largest ones (or the smallest). 
In this second paper we consider the same problem without this restriction which leads to a rather different analysis.
We introduce a new ensemble which is related, but not equivalent, to the ``thinned ensembles'' introduced by Bohigas and Pato. 
% and studied recently in the mathematical literature.
This question is motivated by the study of partial sums of proper time delays in chaotic quantum dots, which are characteristic times of the scattering process.
Using the Coulomb gas technique, we derive the large deviation function for $\tilde{L}$.
Large deviations of linear statistics $L=\sum_{i=1}^N f(\lambda_i)$ are usually dominated by the energy of the Coulomb gas, which scales as $\sim N^2$, implying that the relative fluctuations are of order $1/N$.
For the truncated linear statistics considered here, there is a whole region (including the typical fluctuations region), where the energy of the Coulomb gas is frozen and the large deviation function is purely controlled by an entropic effect. 
Because the entropy scales as $\sim N$, the relative fluctuations are of order $1/\sqrt{N}$.
Our analysis relies on the mapping on a problem of $p$ fictitious non-interacting fermions in $N$ energy levels, which can exhibit both positive and negative effective (absolute) temperatures. 
We determine the large deviation function characterizing the distribution of the truncated linear statistics, and show that, for the case considered here ($f(\lambda)=1/\lambda$), the corresponding phase diagram is separated in three different phases.
\end{abstract}

%\pacs{05.60.Gg ; 03.65.Nk ; 05.45.Mt}

%\pacs{73.20.Fz}{Weak or Anderson localisation}

% 02.50.-r Probability theory, stochastic processes, and statistics
% 02.50.Cw 	Probability theory 
% 02.50.Ey 	Stochastic processes 
% 03.65.Nk    Scattering theory 
% 05.10.Gg 	Stochastic analysis methods (Fokker-Planck, Langevin, etc.) 
% 05.40.-a 	Fluctuation phenomena, random processes, noise, and
% 05.40.Jc Brownian motion
% 05.45.Mt    Quantum chaos ; semiclassical methods 
% 05.60.Gg    Quantum transport
% 05.60.-k 	Transport processes
% 05.70.Np  Interface and surface thermodynamics
%72.   Electronic transport in condensed matter
% 72.10.-d   Theory of electronic transport; scattering mechanisms
% 72.10.Bg   General formulation of transport theory 
% 72.15.Rn Localization effects (Anderson or weak localization)

%73.   Electronic structure and electrical properties of surfaces, interfaces,
%      thin films, and low-dimensional structures 
%73.23.-b     Electronic transport in mesoscopic systems 
%73.20.Fz Weak or Anderson localization

\vspace{0.25cm}


%%%%%%%%%%%%%%%%%%%%%%%%%%%%%%%%%%%%%%%%%%%%%%%%%%%%%%%%%%%%%%%%%%%%%%%%%%%%%%%%%%%%%%%%%%
%%%%%%%%%%%%%%%%%%%%%%%%%%%%%%%%%%%%%%%%%%%%%%%%%%%%%%%%%%%%%%%%%%%%%%%%%%%%%%%%%%%%%%%%%%

\section{Introduction}
\label{sec:Introduction}

The study of linear statistics of eigenvalues of random matrices has played a major role for the applications of random matrix theory to physical problems, like quantum transport \cite{Bee97,KhoSavSom09,MelKum04,VivMajBoh08,VivMajBoh10}, quantum entanglement \cite{PFPPS10,NadMajVer10,NadMajVer11}, etc, as it was emphasized in our recent papers \cite{GraMajTex17,GraTex16b}.
Whereas previous work has considered \textit{unrestricted} sums of the form $L=\sum_{i=1}^Nf(\lambda_i)$, where $\{\lambda_1,\cdots,\lambda_N\}$ is the spectrum of eigenvalues of a random matrix, the question of \textit{truncated} linear statistics (TLS) of the form 
\begin{equation}
  \label{eq:firstEquationTLS}
   \tilde{L}=\sum_{i=1}^\Nt f(\lambda_i)
   \hspace{0.5cm}\mbox{with}\hspace{0.5cm}   
   \Nt<N
\end{equation}
was recently raised in our previous paper \cite{GraMajTex17}. 
There, we have considered the case where the eigenvalues contributing to the partial sum are further constrained to be the $p$ largest (or smallest) ones,
which corresponds to interpolating between two well studied types of problems~: 
for $\Nt=N$,  the statistical analysis of the linear statistics $L=\sum_{i=1}^Nf(\lambda_i)$ \cite{Bee97,CunFacViv16,GraTex15,GraTex16b,KhoSavSom09,NadMajVer10,NadMajVer11,TexMaj13,VivMajBoh08,VivMajBoh10} (and many other references), 
and for $\Nt=1$, the study of the distribution of the largest (or smallest) eigenvalue \cite{BorEynMajNad11,DeaMaj06,DeaMaj08,MajVer09,MajSchVilViv13,MajSch14,TraWid94,TraWid96,VivMajBoh07}.

The aim of the present paper is to analyse the statistical properties of TLS \eqref{eq:firstEquationTLS} \textit{in the absence of further restriction} concerning the ordering of the $p<N$ eigenvalues~; 
we will see that lifting this restriction has strong implications on the analysis and the results. 

A related question was introduced by Bohigas and Pato \cite{BohPat06} who studied the effect of removing  randomly chosen eigenvalues in an invariant matrix ensemble, leading to consider a fraction of the initial eigenvalues.
Initially motivated to model the transition from the random matrix spectrum (correlated eigenvalues) to the Poisson spectrum (uncorrelated eigenvalues), this question has shown a renewed interest recently, under the name of ``thinned ensembles'' \cite{BerDui16,ChaCla16,Lam16}.
Here we study the linear statistics in an ensemble, which is similar to, but different from the thinned ensembles. 
The relation between the two ensembles is discussed below, see Eqs.~(\ref{eq:DistribLambdaN},\ref{eq:RelationWithThinned},\ref{eq:ThinnedEnsemble}).

For convenience, we rescale the TLS as
\begin{equation}
	s = N^{-\eta} \sum_{i=1}^{\Nt} f(\lambda_i)
	\:,
	\label{eq:defTrlinstat0}
\end{equation}
where the exponent $\eta$ is chosen such that $s$ remains of order $N^0$ as $N \to \infty$. 
The precise value of $\eta$ depends on the function $f$ and the matrix ensemble under consideration. 
The general scenario presented in the paper is valid for any choice of function $f$. 
Moreover it applies to different matrix ensembles, although we will focus on Wishart matrices (i.e. the Laguerre ensemble of random matrices).
In the sequel, for reasons explained later, we will consider the case $f(\lambda)=1/\lambda$, leading to $\eta=0$.
Denoting by $P_N(\lambda_1, \cdots, \lambda_N)$ the joint probability distribution function for the eigenvalues (for the Laguerre ensemble, Eq.~\eqref{eq:jpdfGamma} below), we can write the 
distribution of $s$ as~:
\begin{equation}
	\P(s) =  \int \dd \lambda_1 \int \dd \lambda_2 \cdots
			\int \dd \lambda_N \, P_N(\lambda_1, \cdots, \lambda_N)
			\:
		 \delta \bigg(
			s - N^{-\eta} \sum_{i=1}^{\Nt}  f(\lambda_i)
		\bigg)
		\:.
	\label{eq:DefPns0}
\end{equation}
Here, $P_N(\lambda_1, \cdots, \lambda_N)$ is a symmetric function of its $N$ arguments.
% (it could also be another distribution related to a different matrix ensemble)
Due to the restriction to a fraction of the eigenvalues in \eqref{eq:defTrlinstat0}, the only avalaible approaches seem to be the orthogonal polynomial technique, which provides naturally the $p$-point correlation functions, and the Coulomb gas method. 
However the former is restricted to the unitary class and does not permit a simple analysis of the large deviations.
For this reason, we will use the Coulomb gas technique.
Because the method deals with the eigenvalue density, the restriction to a fraction of eigenvalues requires to introduce ``occupation numbers'' $\{\m_i\}_{i=1,\cdots,N}$ and write the TLS as~:
%It is required to order the eigenvalues and introduce ``occupation numbers'' $\m_i$ in order to apply the Coulomb gas method.   $s$ can be rewritten as~:
\begin{equation}
	s = N^{-\eta} \sum_{i=1}^{N} \m_i f(\lambda_i)
	\:,
	\hspace{0.5cm}
	\lambda_1 > \lambda_2 > \cdots > \lambda_N
	\:,
	\label{eq:defTrlinstat}
\end{equation}
where 
%$\m_i \in \lbrace 0, 1 \rbrace$. 
$\m_i=1$ if the eigenvalue $\lambda_i$ contributes to the sum and $\m_i = 0$ otherwise. 
Note that subset sums $\sum_i\m_i\lambda_i$ of eigenvalues of certain random matrices recently appeared in the spectral analysis of fermionic systems~\cite{CunMalMez16}.
Since the sum should contain $\Nt$ eigenvalues, the $N!/\big[\Nt!(N-\Nt)!\big]$ acceptable configurations $\lbrace \m_i \rbrace$ verify 
\begin{equation}
  \sum_{i=1}^{N} \m_i = \Nt
  \:.
\end{equation}
Recently, the case where the summation is restricted to the $\Nt$ largest (or smallest) eigenvalues was considered in Ref.~\cite{GraMajTex17}. This corresponds to considering the configuration $\m_1 = \cdots = \m_{\Nt} = 1$ and $\m_{\Nt+1} = \cdots = \m_{N} = 0$.
Here, we consider the same problem in the absence of this restriction, which leads us to introduce a new ensemble described by the joint distribution for the two sets of random numbers
\begin{equation}
  \label{eq:DistribLambdaN}
  \boxed{
  \mathscr{P}_{N,\Nt}(\{\lambda_i\},\{\m_i\})
  = p!(N-p)!\, P_N(\lambda_1, \cdots, \lambda_N)\,\mathbf{1}_{\lambda_1>\cdots>\lambda_N}\,
  \delta_{\Nt,\sum_i\m_i} 
  }
\end{equation}
where $\mathbf{1}_{\lambda_1>\cdots>\lambda_N}=\prod_{i=1}^{N-1}\heaviside(\lambda_{i}-\lambda_{i+1})$ and $\heaviside(x)$ the Heaviside step function.
Eq.~\eqref{eq:DistribLambdaN} is obviously normalised when integrated over all $\lambda_i$'s and with summation over the occupations.
We can now establish the precise relation with the distribution describing the ``thinned'' random matrix ensembles \cite{BerDui16,BohPat06,ChaCla16,Lam16}, which is obtained by relaxing the constraint on the number $p$. 
Denoting $B_N(p)=\begin{pmatrix}N\\p\end{pmatrix}\kappa^p(1-\kappa)^{N-p}$ the binomial distribution, i.e. the probability to select $p$ eigenvalues among the $N$, where $\kappa$ is the probability to select an eigenvalue,  we have
\begin{align}
  \label{eq:RelationWithThinned}
  \mathscr{P}_{N,\kappa}^\mathrm{(thinned)}(\{\lambda_i\},\{\m_i\})
  &= 
  \sum_{\Nt=0}^N 
  B_N(p)\,
  \mathscr{P}_{N,\Nt}(\{\lambda_i\},\{\m_i\})
  \\
  \label{eq:ThinnedEnsemble}
  &
  \hspace{-1.25cm}
  = N!\, P_N(\lambda_1, \cdots, \lambda_N)\,\mathbf{1}_{\lambda_1>\cdots>\lambda_N}\,
  \kappa^{\sum_i\m_i}   \, (1-\kappa)^{N-\sum_i\m_i} 
  \:,
\end{align}
($p$ then fluctuates, its average being $\overline{p}=\kappa\,N$).

The distribution we aim to determine in the paper can be written in terms of \eqref{eq:DistribLambdaN} as 
\begin{align}
	\nonumber
	\P(s) = \sum_{ \lbrace \m_i \rbrace }
		 \int \dd \lambda_1 \int^{\lambda_1} \dd \lambda_2
		& \cdots
			\int^{\lambda_{N-1}} \dd \lambda_N \,  
			\mathscr{P}_{N,\Nt}(\{\lambda_i\},\{\m_i\})
		\\ \times
		& \delta \bigg(
			s - N^{-\eta} \sum_{i=1}^{N} \m_i f(\lambda_i)
		\bigg)
		\:.		
	\label{eq:DefPns}
\end{align}
where $\kappa=\Nt/N$ in the subscript of $\P(s)$.

\begin{figure}[!ht]
	\centering
	\includegraphics[width=0.95\textwidth]{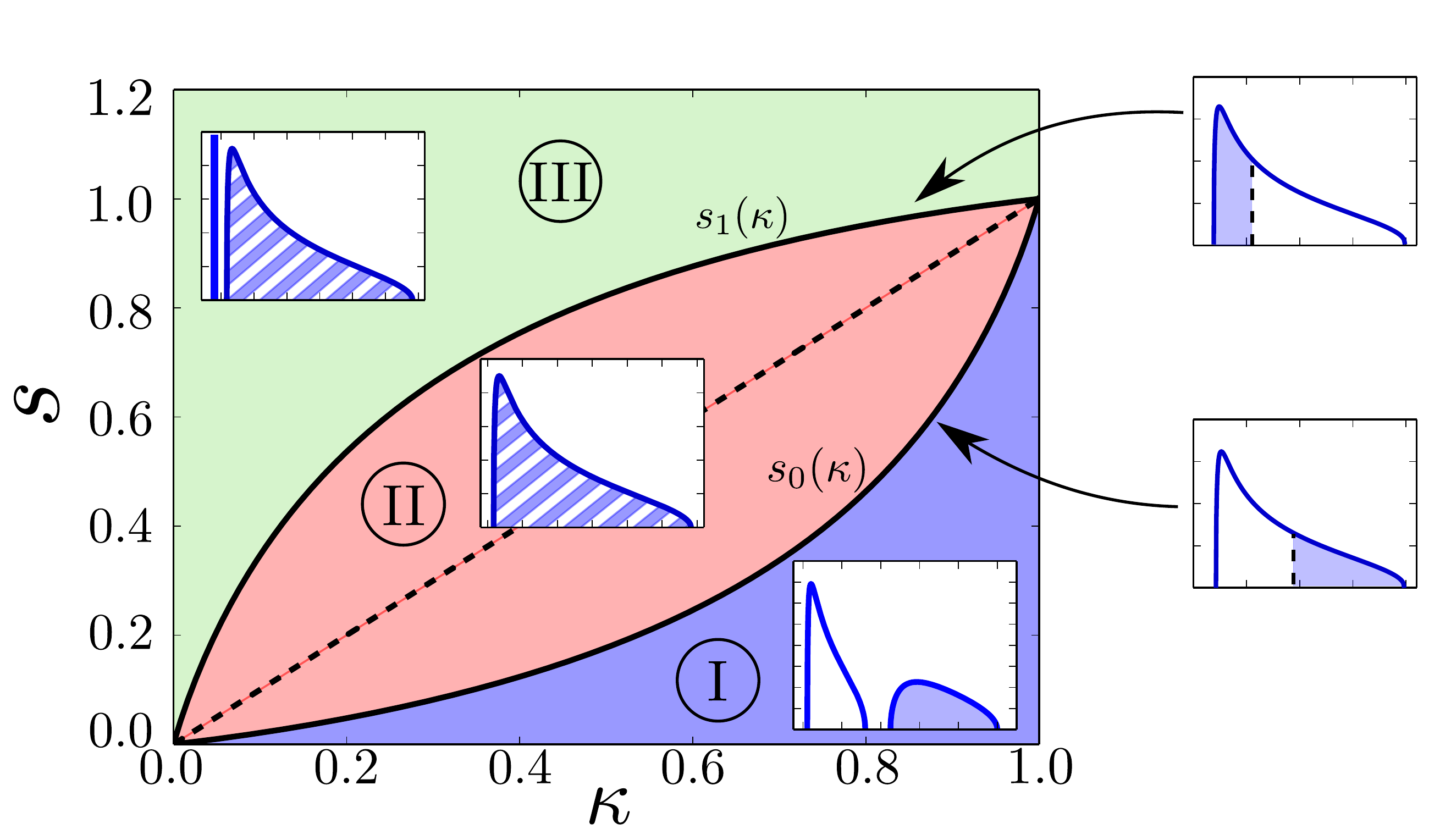}
	\caption{\it 
	  Phase diagram in the $(\kappa,s)$ plane, 
	  where $\kappa=\Nt/N$ and $s=N^{-\eta}\sum_{i=1}^{\Nt} f(\lambda_i)$ 
	  (here for the case $f(\lambda)= 1/\lambda$ in the Laguerre ensemble, thus with $\eta=0$). 
	  Insets show %the shape of 
	  the corresponding optimal ``charge'' density profiles.
	  Shaded part of the density (insets on the right) shows the fraction of (largest or smallest) $\Nt$ eigenvalues contributing to $s$.
	  Dashed profiles (Phases II and III) mean fluctuations in the ``occupations'' $\{\m_i\}$ (cf. text).
	  In the central part of the diagram (Phase II),  
	  the distribution is dominated by entropy, 
	  while in Phases I and III it is dominated by the energy.  
	  }
	\label{fig:PhDiag}
\end{figure}

\subsection{Main results }

\begin{figure}[!ht]
	\centering
	\includegraphics[width=0.8\textwidth]{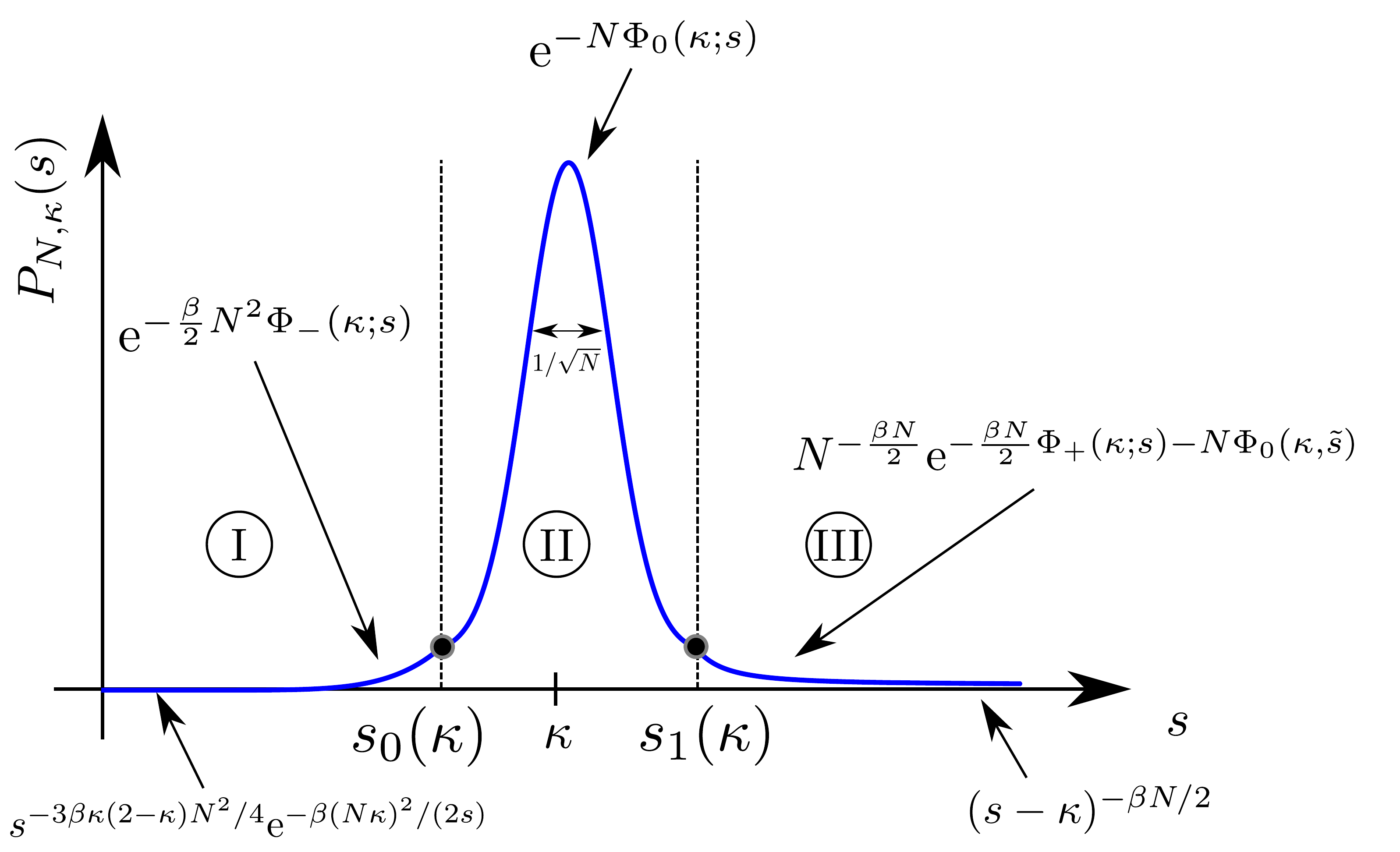}
	\caption{\it 
	 Sketch of the distribution $\P(s)$ obtained in the paper, Eq.~\eqref{eq:SummarisePN}.
	  }
	\label{fig:SchemaDistr}
\end{figure}

Considering the Laguerre ensemble, Eq.~\eqref{eq:jpdfGamma} below, with $f(\lambda)=1/\lambda$, we have obtained the distribution $\P (s)$ in the limit of large matrix size, $N\to\infty$, 
when both $\kappa$ and $1-\kappa$ are of order one~; in particular this excludes the case $\kappa\sim1/N$ which is not covered by our approach (the limit $\kappa\to1$ was also identified as a singular limit in Ref.~\cite{GraMajTex17}).
In this limit the multiple integral \eqref{eq:DefPns} is dominated by the optimal configuration for the two sets of  random numbers. 
As the two parameters $\kappa$ and $s$ are tuned, we obtain different types of optimal configurations, which are interpreted as different ``phases''.
The phase diagram can be drawn in the $(\kappa,s)$ plane, where the physical available phases are located in the stripe $[0,1]\times[0,\infty[$. 
As one increases $s$ for fixed $\kappa$, one encounters three phases described below, the phase transitions taking places at $s=s_0(\kappa)$ and $s=s_1(\kappa)$, which defines two lines in the $(\kappa,s)$ plane (see also Fig.~\ref{fig:PhDiag}): 
\begin{itemize}
\item 
   Phase I~:
   the optimal configuration $\lbrace \lambda_i \rbrace$, that dominates the integrals \eqref{eq:DefPns}, varies with $(\kappa,s)$ while the occupations $\lbrace \m_i \rbrace$ remain frozen.
   This case is conveniently analysed within the conventional Coulomb gas interpretation where the eigenvalues are considered as the positions of a one-dimensional gas of $N$ particles with logarithmic interactions.
	  
\item 
   Phase II~:
   $\lbrace \lambda_i \rbrace$ are frozen while $\lbrace \m_i \rbrace$ fluctuate.
	To get some insight, we interpret the problem as a gas of $p$ fictitious fermions occupying $N$ fixed ``energy levels'' $\varepsilon_i=f(\lambda_i)$'s. % (the fermionic nature arise from the fact that the occupations are $n_i=0$ or $1$). 
	This phase includes the typical fluctuations (the dashed line in Fig.~\ref{fig:PhDiag} indicates where $\P (s)$ is maximum).

\item
   Phase III~:
   the last phase is described in terms of a
   mixed picture where one single eigenvalue splits off the bulk, while the density of the remaining eigenvalues is frozen and the occupation numbers fluctuate (they freeze as $s\to\infty$).	
\end{itemize}
The different phases are related to the following behaviours for the distribution of the TLS~\footnote{Expressions of the type $\P(s) \underset{N\to\infty}{\sim} \exp(-N^q \Phi)$ must be understood as
$\lim_{N \to \infty} (-1/N^q) \ln \P(s) = \Phi$.
}~:
\begin{align}
  \label{eq:SummarisePN}
  &\P (s)  
  \\
  \nonumber
  &\underset{N \to \infty}{\sim}\begin{cases}
    \e^{- N\,C_\kappa}
    \exp \left\lbrace - \frac{\beta N^2}{2} \Phi_-(\kappa;s)  \right\rbrace
    & \text{for }    s < s_0(\kappa) 
    \\[0.2cm]
   \exp \left\lbrace - N \Phi_0(\kappa;s) \right\rbrace	
	\hspace{0.5cm}
	& \text{for }    s_0(\kappa)<s<s_1(\kappa)
	\\[0.2cm]
	N^{-\frac{\beta N}{2}}
	\exp \left\lbrace -
	   N\left[ \frac{\beta }{2}\Phi_+(\kappa;s) + \Phi_0(\kappa; \tilde{s}(s)) \right] \right\rbrace
	& \text{for }     s > s_1(\kappa)
  \end{cases}
\end{align}
where $\tilde{s}(s)$ smoothly interpolates between $s_1(\kappa)$ and $\kappa$ (it will be studied in Section~\ref{sec:LDFlarge}).
The large deviation function in the last expression combines two different functions with different origins (energy \textit{versus} entropy).
A sketch of the distribution is represented in Fig.~\ref{fig:SchemaDistr}.
The different scalings with $N$ in the exponential arise from the fact that entropy $\sim N$ dominates in Phase II,  including the typical fluctuations, while energy $\sim N^2$ dominates in Phase I  (the scaling for Phase III is due to a subleading contribution to the energy).
The scaling in Phase II implies in particular that the fluctuations of the TLS scale as $\sim1/\sqrt{N}$ (and are independent of the symmetry index $\beta$).
The limiting behaviours of the large deviation functions for the three regimes are~:
\begin{align}
	&\Phi_-(\kappa;s) \simeq
	\begin{cases}
		\displaystyle
		\frac{\kappa^2}{s} + \frac{3 \kappa(2-\kappa)}{2} \ln s
		+ \mathrm{cste}
		& \text{as } s \to 0
		\\[0.2cm]
		\displaystyle
		\omega_\kappa\, (s - s_0(\kappa))^2
		& \text{as } s \to s_0(\kappa)^-
	\end{cases}
	\\
	&\Phi_0(\kappa;s) \simeq
	\begin{cases}
		%\displaystyle
		C_{\kappa}
		- \cp \sqrt{s - s_{0}(\kappa) }
		& \text{as } s \to s_0(\kappa)^+
		\\[0.2cm]
		\displaystyle
		\frac{(s-\kappa)^2}{2 \kappa(1-\kappa)}
		& \text{as } s \to \kappa
		\\[0.2cm]
		%\displaystyle
		C_{\kappa}
		- \cpt \sqrt{ s_{1}(\kappa) - s }
		& \text{as } s \to s_1(\kappa)^-
	\end{cases}
	\\
	&\Phi_+(\kappa;s) + \frac{2}{\beta} \Phi_0(\kappa;\tilde{s}(s)) \simeq
	\begin{cases}
		\displaystyle
		 \ln(s - s_1(\kappa) )  + \mathrm{cste}
		& \text{as } s \to s_1(\kappa)^+
		\\[0.2cm]
		\displaystyle
		 \ln (s - \kappa)
		 + \mathrm{cste}
		& \text{as } s \to +\infty
	\end{cases}
\end{align}
where $\omega_\kappa$, $C_{\kappa}=-\kappa \ln \kappa - (1-\kappa) \ln (1-\kappa)$, $\cp$ and $\cpt$ are positive constants.

\subsection{Plan}

Section~\ref{sec:Motivation} exposes the motivations which have led us to the study of the distribution of the TLS for $f(\lambda)=1/\lambda$.
In Section~\ref{sec:Coulombgas}, we will describe Phase I, which will be analysed within the usual Coulomb gas picture. We will introduce the notations. 
The derivation of the results has been made as short as possible, as the problem is similar to the one studied in our previous article~\cite{GraMajTex17}, although the case considered here is more complicated (in Ref.~\cite{GraMajTex17}, we have made use of simplifications arising from the specific choice $f(\lambda)=\sqrt{\lambda}$).
Section~\ref{sec:TypFluct} will discuss Phase II, analysed within the picture of fictitious non interacting fermions.
Phase III is studied in Section~\ref{sec:LDFlarge}.
The paper is closed by some concluding remarks.
%The distribution of the partial sum $\sum_{i=1}^p\WSm_{ii}$ is studied in Appendix~\ref{app:Injectance}.

%%%%%%%%%%%%%%%%%%%%%%%%%%%%%%%%%%%%%%%%%%%%%%%%%%%%%%%%%%%%%%%%%%%%%%%%%%%%%%%%%%%%%%%%%%
%%%%%%%%%%%%%%%%%%%%%%%%%%%%%%%%%%%%%%%%%%%%%%%%%%%%%%%%%%%%%%%%%%%%%%%%%%%%%%%%%%%%%%%%%%

\section{Motivations~: proper time delays \textit{versus} diagonal elements of the Wigner-Smith matrix in chaotic scattering}
\label{sec:Motivation}

\subsection{Chaotic scattering in quantum dots and random matrices}
\label{sec:QuantScatt}

\begin{figure}[!ht]
\centering
\includegraphics[width=0.4\textwidth]{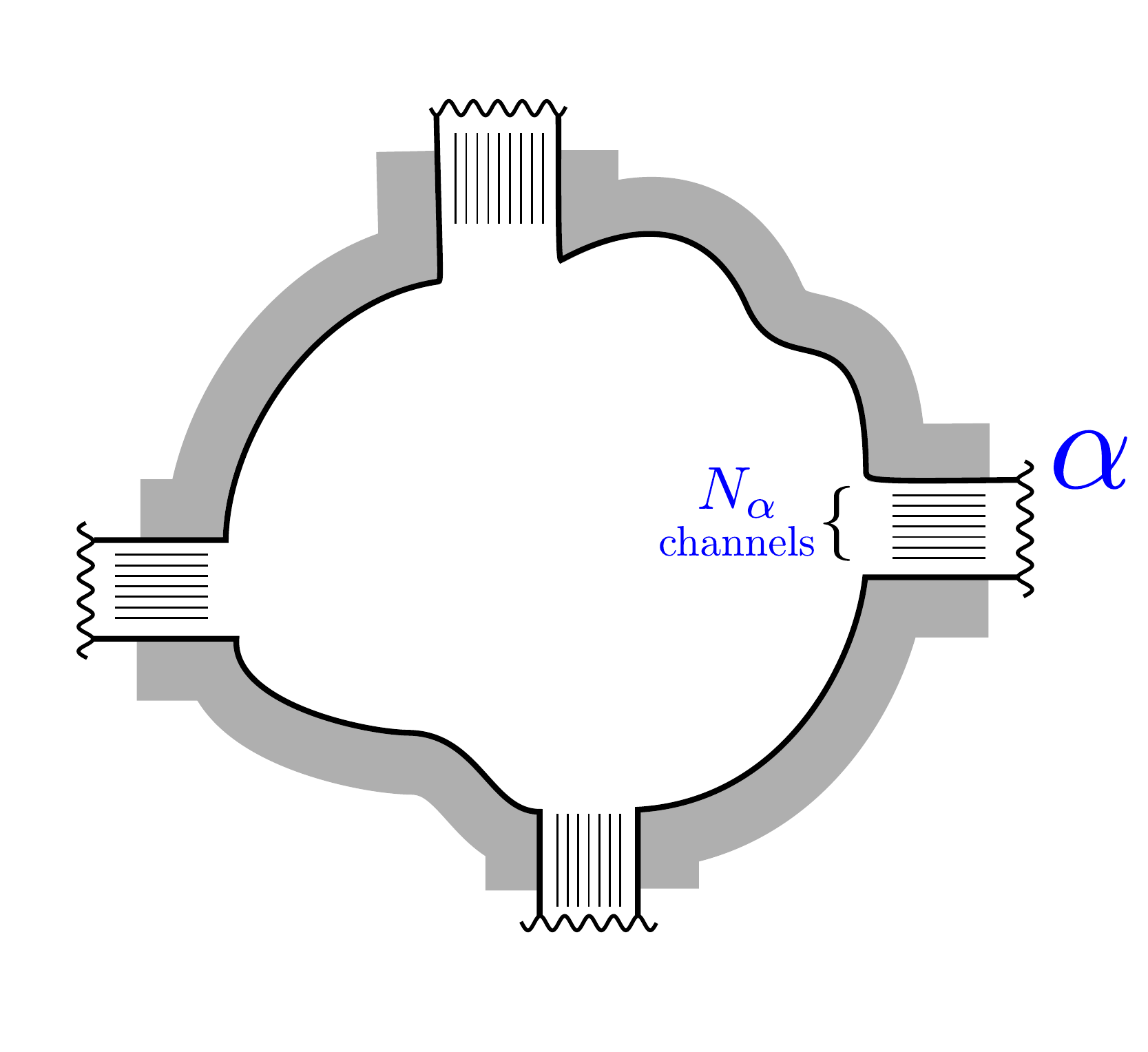}
\caption{\it A chaotic quantum dot with four perfect contacts. Contact $\alpha$ has $N_\alpha$ open channels.}
\label{fig:qd}
\end{figure}

The problem studied in this article is motivated by the statistical analysis of partial sums of proper time delays in chaotic scattering.
Proper time delays are characteristic times capturing temporal aspects of quantum scattering (for a review on time delays, cf.~\cite{CarNus02}~; aspects related to chaotic scattering are reviewed in \cite{FyoSom97,Kot05,KuiSavSie14,Tex16}).
For the sake of concreteness, let us consider a quantum dot (QD) connected to external contacts playing the role of wave guides through which electrons can be injected (Fig.~\ref{fig:qd}). 
An electronic wave of energy $\varepsilon$ injected in the conducting channel $a$ is scattered in channel $b$ with probability amplitude given by the matrix element $\Sm_{ba}(\varepsilon)$ of the $N\times N$ on-shell scattering matrix $\Sm(\varepsilon)$.
Probing the energy structure of  $\Sm(\varepsilon)$ allows one to define several characteristic times.
Ref.~\cite{Tex16} has emphasized the difference between the \textit{partial} time delays $\{\tilde{\tau}_i\}$, which are derivatives of the scattering phase shifts, and the \textit{proper} time delays $\{\tau_i\}$, which are eigenvalues of the Wigner-Smith matrix  
\begin{equation}
  \WSm=-\I \Sm^\dagger\partial_\varepsilon \Sm
  \:.
\end{equation}
In other words, these two sets of characteristic times are obtained by performing derivation and diagonalisation in different orders.
They however satisfy the sum rule $\sum_i\tilde\tau_i=\sum_i\tau_i$.

The Wigner-Smith matrix is a central concept allowing to characterize the amount of injected charge in the QD, which can be understood from Krein-Friedel type formulae relating the scattering matrix to the density of states.
This observation has played a major role in B\"uttiker's quantum transport theory for coherent devices beyond DC linear transport~: non-linear transport \cite{ChrBut96a}, AC transport \cite{ButPreTho93,ChrBut96,GraTex15}, pumping, frequency dependent noise, etc (see \cite{Tex16} for a review and further references).
An important observation is that, in order to address the out-of-equilibrium situation, the contribution of the electrons from a given contact must be identified.
This has led to the concept of \textit{injectance},
$\overline{\nu}_\alpha=(2\pi)^{-1}\sum_{i\in\mathrm{contact}\:\alpha}\WSm_{ii}$,
measuring the contribution to the density of states (DoS) inside the device of the scattering states incoming from the channels associated with the contact $\alpha$~;
hence the total DoS is $\sum_\alpha\overline{\nu}_\alpha=(2\pi)^{-1}\tr{\WSm}$, which is related to the Wigner time delay~:
\begin{equation}
  \label{eq:DoSSumRule}
  \tau_W %= \frac{1}{N}\tr{\WSm} 
  = \frac{1}{N}\sum_i \WSm_{ii} = \frac{1}{N}\sum_i\tau_i
  \:.
\end{equation}
The symmetric concept of \textit{emittance}, 
$\underline{\nu}_\alpha=(2\pi)^{-1}\sum_{i\in\mathrm{contact}\:\alpha}[\Sm\WSm\Sm^\dagger]_{ii}$, 
was also introduced in order to allow for a postselection of the contact,
whereas the injectance $\overline{\nu}_\alpha$ corresponds to the preselection (see Ref.~\cite{GasChrBut96} and the review \cite{Tex16}).

In systems with complex dynamics, such as chaotic quantum dots (Fig.~\ref{fig:qd}), the statistical approach is the most efficient, which leads to assume that the scattering matrix $\Sm$ belongs to a random matrix ensemble~\cite{Bee97,MelKum04}. 
In chaotic QDs with perfect contacts, it is quite natural to choose the uniform distribution over the unitary group, up to some additional constraints related to the symmetry (time reversal symmetry and/or spin rotation symmetry)~: this corresponds to the so-called circular matrix ensembles.
This approach however does not provide any information on the energy structure of the scattering matrix, what is required in order to characterize the statistical properties of the matrix $\WSm$.
An ``alternative stochastic approach'' has been introduced in \cite{BroBut97} and applied in \cite{BroFraBee97,BroFraBee99} where the Wigner-Smith matrix distribution was shown to be related to the Laguerre ensemble of random matrices~\cite{BroFraBee97,BroFraBee99}.~\footnote{
  Refs.~\cite{BroFraBee97,BroFraBee99} has introduced the symmetrized Wigner-Smith matrix $\WSm_s=-\I\Sm^{-1/2}\,\partial_\varepsilon\Sm\,\Sm^{-1/2}$, with the same spectrum of eigenvalues than $\WSm$.
  The precise statement of these references is that $1/\WSm_s$ is a Wishart matrix.
  Because $\Sm$ and $\WSm$ are only independent in the unitary case, $1/\WSm$ is a Wishart matrix only in this case, strictly speaking.
  However its eigenvalues are always given by the Laguerre distribution~\eqref{eq:jpdfGamma}.
}
The joint distribution of the inverse of the proper times $\lambda_i=\Ht/\tau_i$ is
\begin{equation}
	P_N(\lambda_1, \cdots, \lambda_N) \propto
		\prod_{i<j} \abs{\lambda_i - \lambda_j}^\beta
		\prod_{n=1}^N \lambda_n^{\beta N/2} \e^{-\beta \lambda_n/2}
		\:,
	\label{eq:jpdfGamma}
\end{equation}
where $\beta\in\{1,\,2,\,4\}$ is the Dyson index corresponding to orthogonal, unitary and symplectic classes.
The time scale denotes the Heisenberg time $\Ht=2\pi/\Delta$, where $\Delta$ is the mean level spacing (or mean resonance spacing).
The DoS interpretation thus leads to $(2\pi)^{-1}\smean{\tr{\WSm}}=1/\Delta$.
Combined with the sum rule \eqref{eq:DoSSumRule}, we have 
\begin{equation}
  \smean{\tau_W}=\smean{\WSm_{ii}}=\smean{\tau_i}= \frac{\Ht}{N}
  \:.
\end{equation}
This scale is also known as the dwell time and measures the average time spent by a wave packet inside the quantum dot.
Later we will set $\Ht=1$ for simplicity.
The variance and the correlations of proper times are also known (see updated arXiv version of \cite{Tex16} and references therein)~\cite{Sav16}~:
\begin{align}
  \label{eq:VarianceTaua}
	\Var(\tau_i) &= \frac{N[\beta(N-1)+2] + 2}{N^2(N+1)(\beta N - 2)}
	\:,
	\\
  \label{eq:CovarianceTaua}
	\Cov(\tau_i, \tau_j) &= - \frac{1}{N^2(N+1)} 
	\hspace{1cm}\mbox{for } i\neq j
	\:.
\end{align}
(other references on time delay correlations are \cite{KuiSavSie14,Nov15a,CunMezSimViv16b}, and can also be found in the review papers quoted above).

The joint distribution for the partial time delays $\tilde{\tau}_i$'s and the diagonal matrix elements $\WSm_{ii}$'s are still unknown.
An interesting connection between them was pointed out in \cite{SavFyoSom01} (see also \cite{KuiSavSie14})~:
introducing the unitary matrix $\mathcal{V}$ which diagonalizes the scattering matrix, these authors have obtained the non trivial relation $\tilde{\tau}_i=\big[\mathcal{V}^\dagger \WSm\mathcal{V}\big]_{ii}$.
In the unitary case, $\mathcal{V}$ is expected to be independent of $\WSm$. 
Because $\WSm$ is invariant under unitary transformations in this case, we have $\tilde{\tau}_i=\WSm_{ii}$. 
%(in the unitary case, eigenvalues of $Q$ and eigenvectors, i.e. $\mathcal{U}$, are also independent \cite{BroFraBee99}). 
Let us emphasize few properties of the $\WSm_{ii}$'s, which are important as they are involved in the injectances  which should be accessible experimentally, and emphasize their difference with the proper times.
$\WSm_{ii}$'s and $\tau_i$'s can be related by writing 
$\WSm=\mathcal{U}\,\mathrm{diag}(\tau_1,\,\cdots,\,\tau_N)\,\mathcal{U}^\dagger$  (the unitary matrices $\mathcal{V}$ and $\mathcal{U}$ differ in general), so that 
\begin{equation}
  \label{eq:RelationQiiTau}
  \WSm_{ii} = \sum_{j=1}^N \left|\mathcal{U}_{ij}\right|^2\tau_j
  \:.
\end{equation}
In the unitary case ($\beta=2$) the matrix $\mathcal{U}$ and the eigenvalues are uncorrelated, which allows to go further in the statistical analysis of the $\WSm_{ii}$'s (this is not the case in the orthogonal and symplectic cases~\cite{BroFraBee99}). Thus we will focus on the unitary case for a moment, and will come back to the general case later.
Correlations of the $\mathcal{U}$ matrix elements are controlled by Weingarten functions \cite{Col03}.
Some algebra gives~\footnote{
  We have used
  \begin{align*}
	\int_{U(N)} \dd U\, 
	  U_{i_1 j_1} U_{i_2 j_2} U^*_{k_1 l_1} U^*_{k_2 l_2}
		&= W(N,1^2)
		\left(
			\delta_{i_1 k_1} \delta_{i_2 k_2} \delta_{j_1 l_1} \delta_{j_2 l_2}
			+ \delta_{i_1 k_2} \delta_{i_2 k_1} \delta_{j_1 l_2} \delta_{j_2 l_1}
		\right)
		\\ \nonumber
		& + W(N,2)
		\left(
			\delta_{i_1 k_1} \delta_{i_2 k_2} \delta_{j_1 l_2} \delta_{j_2 l_1}
			+ \delta_{i_1 k_2} \delta_{i_2 k_1} \delta_{j_1 l_1} \delta_{j_2 l_2}
		\right)
		\:,
  \end{align*}
  where the Weingarten functions $W(N,\sigma)$ is a function of the matrix size and the permutation of indices. In particular~:
  $
	W(N,1^2) = \frac{1}{N^2-1}
  $
  and
  $
	W(N,2) = -\frac{1}{N(N^2-1)}
  $.
} 
\begin{equation} 
  \label{eq:CovQii}
  \Cov\left( \WSm_{ii} , \WSm_{jj} \right) = 
  \frac{1}{N(N^2-1)}
  \left(\frac{1}{N} + \delta_{i,j}\right)
  \:.
\end{equation}
In the particular case $i=j$, this result coincides with the one first obtained in \cite{FyoSom97} (note that this reference has also obtained the variance for non ideal leads).
Eq.~\eqref{eq:CovQii}
matches with the covariances of the partial times given in the updated arXiv version of \cite{Tex16},
$\Cov\left( \WSm_{ii} , \WSm_{jj} \right) = \Cov(\tilde\tau_i, \tilde\tau_j)$
(see the discussion in Appendices A and E of \cite{KuiSavSie14}).
Both the scaling with $N$ and the sign of the correlations for $\WSm_{ii}=\tilde{\tau}_i$'s and $\tau_i$'s differ~:
$\Var\big( \WSm_{ii} \big)\simeq(1/N)\,\Var(\tau_i)\simeq1/N^3$ and 
$\Cov\big( \WSm_{ii} , \WSm_{jj} \big)\simeq-(1/N)\,\Cov(\tau_i, \tau_j)\simeq+1/N^4$, 
where the factor $1/N$ can be understood as the contribution of the unitary matrix elements in \eqref{eq:RelationQiiTau}.
%%%%%%%%%%%%%% CUT %%%%%%%%%%%%%%%%%%%
\iffalse
The striking difference between the two quantities is best understood from the relation 
\begin{equation}
  \Var\big( \WSm_{ii} \big) = \frac{1}{N+1}
  \left[
    \frac{1}{N}\Var\big(\sum_i\tau_i\big) + \Var(\tau_i)
  \right]
  \:.
\end{equation}
If $\tau_i$'s were identical and idependent random variables, $\Var\big(\sum_i\tau_i\big)=N\Var(\tau_i)$, the first term would dominate which would lead to ``$\Var\big( \WSm_{ii} \big)\simeq(2/N)\Var(\tau_i)$''. 
Instead, the proper times are strongly anti-correlated, resulting in $\Var\big(\sum_i\tau_i\big)\sim\Var(\tau_i)$ and therefore $\Var\big( \WSm_{ii} \big)\simeq(1/N)\Var(\tau_i)$.
\fi
%%%%%%%%%%%%%% END OF CUT %%%%%%%%%%%%%%%%%%%

\subsection{Partial sums of $\tau_i$'s or $\WSm_{ii}$'s : different scaling with $N$}

These relations lead to important differences bewteen partial sums of $\WSm_{ii}$'s and $\tau_i$'s.
Considering a contact with $p$ conducting channels, the two first cumulants of the related injectance  are
\begin{align}
  \Big\langle \sum_{i=1}^p  \WSm_{ii} \Big\rangle 
  &= \frac{p}{N} \equiv \kappa 
  \\
  \label{eq:VarianceSumQii}
  \Var\Big( \sum_{i=1}^p \WSm_{ii} \Big)
  &= \frac{\kappa(1+\kappa)}{N^2-1}
\end{align}
(the leading order to the second expression was given in \cite{Tex16}, Eq.~85 of the arXiv version~; it was deduced from the results of \cite{BroBut97}).
On the other hand, the fluctuations of the partial sum of proper time delays can be deduced from (\ref{eq:VarianceTaua},\ref{eq:CovarianceTaua}) by writing
$
\Var\big( \sum_{i=1}^p  \tau_{i} \big)
= 
p\, \Var(\tau_i) + p(p-1)\, \Cov(\tau_i,\tau_j)
$, explicitly
\begin{equation}
  \label{eq:VarianceSumTaui}
  \Var\Big( \sum_{i=1}^p  \tau_{i} \Big)
  = \frac{\kappa(1-\kappa)}{N+1} + \frac{2\kappa}{N^2-1}
  \:.
\end{equation}
Comparing \eqref{eq:VarianceSumQii} and \eqref{eq:VarianceSumTaui} leads to two interesting observations~:
\begin{itemize}
\item  For $\kappa<1$, the variance of the injectance  is $\sim(1/N)$ smaller than the variance of partial sum of $\tau_i$'s~:
  $\Var\big( \sum_{i=1}^p\WSm_{ii} \big)\sim(1/N)\,\Var\big( \sum_{i=1}^p  \tau_{i} \big)$.
\item  For $\kappa=1$, they are both equal to $\Var\big(\tr{\WSm}\big)=2/(N^2-1)$ as the two quantities satisfy the sum rule~\eqref{eq:DoSSumRule}. 
\end{itemize}
This is a consequence of the fact that $\Var\big( \sum_{i=1}^p  \tau_{i} \big)$ drops by a factor $1/N$ as $\kappa\to1$.
In other words, $\Var\big( \sum_{i=1}^p  \tau_{i} \big)/\smean{ \sum_{i=1}^p  \tau_{i} }^2$ crosses over from $\simeq(1-\kappa)/(\kappa N)\sim1/p$, as in the case of $p$ independent variables, to $\simeq2/N^2$, characteristic of the $N$ highly correlated eigenvalues of random matrices. 
A similar change of scaling has been recently identified in the ``thinned circular ensemble'' in Ref.~\cite{BerDui16}.
These intringuing features have led us to analyse the full distribution of these two partial sums.
The distribution of the injectance $\overline{\nu}=(2\pi)^{-1}\sum_{i=1}^p \WSm_{ii}$ is obtained in Appendix~\ref{app:Injectance} for $\beta=2$. 
This derivation is based on a result of Savin, Fyodorov and Sommers~\cite{SavFyoSom01}, stating that, for $\beta=2$, the distribution of a sub-block of the matrix $\WSm$ is also related to the Laguerre distribution.~\footnote{The more general statement of Ref.~\cite{SavFyoSom01} concerns sub-block of matrix $\WSm_s$, introduced in the previous footnote. The distributions of the two matrices $\WSm$ and $\WSm_s$ coincide for $\beta=2$.}
This makes this study a variant of the Wigner time delay distribution analysis of Ref.~\cite{TexMaj13}~:
the distribution of the partial sum $\sum_{i=1}^p\WSm_{ii}$ is studied in Appendix~\ref{app:Injectance}.
On the other hand, the question of the distribution of the truncated linear statistics
\begin{equation}
  \label{eq:LSofInterest}
  s = \sum_{i=1}^p \tau_i
   = \Ht \sum_{i=1}^p \frac{1}{\lambda_i}  
  \:,
\end{equation}
where $\lambda_i$'s are eigenvalues of a Wishart matrix, Eq.~\eqref{eq:jpdfGamma}, is much more challenging and is the main object of investigation of the present article.

\subsection{Variance for arbitrary symmetry class}

The discussion of the statistical properties of $\WSm_{ii}$'s has led us to restrict ourselves to the unitary case.
However, as we will study the full distribution of $s$ for arbitrary symmetry class, it is useful to express the two cumulants of $s$ for arbitrary $\beta$.
Using Eqs.~(\ref{eq:VarianceTaua},\ref{eq:CovarianceTaua}) we get
\begin{align}
	\label{eq:MeanS}
	\mean{s} &= \sum_{i=1}^\Nt \mean{\tau_i} = \frac{\Nt}{N} = \kappa
	\:,
    \\
	\label{eq:VarSfiniteN}
	\Var(s) 
	%&= \sum_{n=1}^\Nt \Var(\tau_n) + \sum_{n \neq m = 1}^\Nt \Cov(\tau_n, \tau_m)\\
	&= \frac{\beta \Nt N(N - \Nt) + 2 \Nt(\Nt+N)}{N^2(N+1)(\beta N - 2)}
	\\
	\label{eq:VarS}
	&= \frac{\kappa(1-\kappa)}{N}
		+ \frac{4 \kappa}{\beta N^2} \left( 1 - \beta \frac{1-\kappa}{4} \right) + \O(N^{-3})
	\:.
\end{align}
Setting $\kappa = 1$, we check that \eqref{eq:VarSfiniteN} gives the result of Ref.~\cite{MezSim13}~:
\begin{equation}
	\Var(s) 
	= \frac{4}{(N+1)(N\beta-2)} 
	\underset{N \to \infty}{\simeq}
		\frac{4}{\beta N^2}
\end{equation}
(the leading order was obtained earlier in \cite{LehSavSokSom95} for $\beta=1$ and \cite{BroBut97}~; see also \cite{TexMaj13,Tex16}).
Eq.~\eqref{eq:VarS} exhibits another interesting feature as the leading order term is independent of the symmetry class, while the subleading term surviving in the $\kappa\to1$ limit does depend on $\beta$.
The origin of this observation will be clarified in Section~\ref{sec:TypFluct}.

%%%%%%%%%%%%%%%%%%%%%%%%%%%%%%%%%%%%%%%%%%%%%%%%%%%%%%%%%%%%%%%%%%%%%%%%%%%%%%%%%%%%%%%%%%
%%%%%%%%%%%%%%%%%%%%%%%%%%%%%%%%%%%%%%%%%%%%%%%%%%%%%%%%%%%%%%%%%%%%%%%%%%%%%%%%%%%%%%%%%%

\section{Phase I ($s < s_0(\kappa)$)~: Coulomb gas}
\label{sec:Coulombgas}

\subsection{Formalism~: Coulomb gas and occupation numbers}

Let us introduce the general framework used in the paper. For large $N$, it is convenient to rescale the eigenvalues as \cite{GraMajTex17}
\begin{equation}
	\lambda_i = N x_i \:.
\end{equation}
The Coulomb gas method consists in rewriting the joint distribution of eigenvalues (\ref{eq:jpdfGamma}) as a Gibbs measure $\exp(- \beta N^2 \mathcal{E}_\mathrm{gas}[\lbrace x_i \rbrace]/2)$, where
\begin{equation}
	\mathcal{E}_\mathrm{gas}[\lbrace x_i \rbrace] = 
		-\frac{1}{N^2} \sum_{i \neq j} \ln \abs{x_i - x_j}
		+ \frac{1}{N} \sum_{i=1}^N (x_i - \ln x_i)
	\label{eq:Ediscr}
\end{equation}
is the energy of a gas of particles on a line, at positions $\lbrace x_i \rbrace$, interacting with logarithmic repulsion and trapped in the external potential $V(x) = x - \ln x$.
We want to compute the distribution of
\begin{equation}
	s = \frac{1}{N} \sum_{i=1}^N \m_i f(x_i)
	\:,
	\hspace{0.5cm}
	x_1 > x_2 > \cdots > x_N
	\:,
	\label{eq:DefSRescaled}
\end{equation}
for any given function $f$. This is given by Eq.~(\ref{eq:DefPns}), which rewrites~:
\begin{align}
	\label{eq:PnsRatioMultInt}
	&\P(s) = \\ 
	&\frac{ \displaystyle
		\sum_{ \lbrace \m_i \rbrace }
		 \int_0^\infty \hspace{-0.25cm} \dd x_1 \int_0^{x_1} \hspace{-0.25cm}\dd x_2 \cdots
			\int_0^{x_{N-1}}\hspace{-0.25cm} \dd x_N \,
			\e^{-\frac{\beta N^2}{2} \mathcal{E}_\mathrm{gas}[\lbrace x_i \rbrace] }
		\delta \left(
			s - \frac{1}{N} \sum_{i=1}^{N} \m_i f(x_i)
		\right)
		\,
		\ConstN
	}
	{ \displaystyle
		\sum_{ \lbrace \m_i \rbrace }
		 \int_0^\infty \hspace{-0.25cm}\dd x_1 \int_0^{x_1} \hspace{-0.25cm}\dd x_2 \cdots
			\int_0^{x_{N-1}}\hspace{-0.25cm} \dd x_N \,
			\e^{-\frac{\beta N^2}{2} \mathcal{E}_\mathrm{gas}[\lbrace x_i \rbrace] }
			\,
			\ConstN
	}
	\nonumber
		\:.
\end{align}

Since we are interested in partial sums of proper time delays $\tau_i = 1/\lambda_i = 1/(N x_i)$, we will consider the case 
$$
	f(x) = \frac{1}{x}
	\:.
$$
Therefore, Eq.~(\ref{eq:DefSRescaled}) gives $s = N^{-1} \sum_i \m_i / x_i$. The limit $s \to 0$ corresponds to $x_i \to \infty$, for all $\Nt$ eigenvalues such that $\m_i = 1$. The $N - \Nt$ others which do not contribute to the sum are not constrained, so they remain of order~1. Thus, in this case, the $\Nt$ eigenvalues in the sum are the largest. This corresponds to
\begin{equation}
	\m_1 = \cdots = \m_{\Nt} = 1 \: ,
	\hspace{0.5cm}
	\m_{\Nt+1} = \cdots = \m_{N} = 0 \:.
	\label{eq:niFrozen}
\end{equation}
Eq.~(\ref{eq:DefSRescaled}) reduces to~:
\begin{align}
	s = \frac{1}{N} \sum_{i=1}^{\Nt} f(x_i)
	\:,
	\hspace{0.5cm}
	x_1 > x_2 > \cdots > x_N
		\:.
	\label{eq:sLargest}	
\end{align}
This problem was analysed in detail in \cite{GraMajTex17}, for some specific choices of function $f$. The general case is discussed in the appendix of this paper. Here we will only sketch briefly the method in order to be self-contained.

In the remainder of this section, we will focus on the situation where the occupation numbers $\lbrace \m_i \rbrace$ are frozen according to (\ref{eq:niFrozen}). This will be the case as long as $s < s_0(\kappa)$.
In Sections~\ref{sec:TypFluct} and~\ref{sec:LDFlarge}, we will study the case $s > s_0(\kappa)$ where the $\lbrace \m_i \rbrace$'s are allowed to fluctuate.

\subsection{Functional integral formulation}

Let us introduce the empirical density of eigenvalues
\begin{equation}
	\rho(x) = \frac{1}{N} \sum_{i=1}^N \delta(x - x_i)
	\:.
	\label{eq:defEmpDens}
\end{equation}
In terms of this density, (\ref{eq:sLargest}) reads~:
\begin{equation}
	s = \int_c \rho(x)f(x) \dd x \:,
	\label{eq:ConstrScoulomb}
\end{equation}
where $c \sim x_{\Nt}$ is a lower bound ensuring that only the $\Nt$ largest eigenvalues contribute to the integral. It is fixed by imposing
\begin{equation}
	\int_c \rho(x) \dd x = \frac{\Nt}{N} = \kappa
	\:.
	\label{eq:ConstrKcoulomb}
\end{equation}
We emphasize that it is only possible to express $s$ in terms of a simple integral over the density of eigenvalues $\rho(x)$ in the case where only the largest (or the smallest) eigenvalues contribute to $s$, corresponding to large deviations.
In the limit $N \to \infty$, it is possible to rewrite the multiple integrals in Eq.~(\ref{eq:PnsRatioMultInt}) as functional integrals over the density, leading to the measure~\cite{DysonI,DeaMaj08}~:
\begin{equation}
	\e^{-\frac{\beta N^2}{2} \mathcal{E}_\mathrm{gas}[\lbrace x_i \rbrace] }
	\dd x_1 \cdots \dd x_N
		\rightarrow
	\e^{- \frac{\beta N^2}{2} \mathscr{E}[\rho] + N (1 - \frac{\beta}{2}) \mathscr{S}[\rho]}
	\D \rho
	\:,
	\label{eq:MeasurePathInt}
\end{equation}
with the energy
\begin{equation}
	\mathscr{E}[\rho] = - \int \dd x \int \dd y \: \ln \abs{x-y}
		+ \int \dd x \: \rho(x)(x - \ln x)
	\label{eq:Erho}
\end{equation}
and the entropy
\begin{equation}
	\mathscr{S}[\rho] = -\int \dd x \: \rho(x) \ln \rho(x)
	\:.
	\label{eq:Srho}
\end{equation}
The energy $\mathscr{E}[\rho]$ is obtained by rewriting (\ref{eq:Ediscr}) in terms of the density (\ref{eq:defEmpDens}). The diagonal terms $i=j$ which are not present in (\ref{eq:Ediscr}) are removed by adding the term $-\big[{\beta}/{(2N)}\big] \mathscr{S}[\rho]$ \cite{DysonII,DeaMaj08}. The other entropic term $({1}/{N}) \mathscr{S}[\rho]$ comes from the loss of information when describing the set $\lbrace x_i \rbrace$ by the density $\rho(x)$.

Since we are interested in the limit $N \to \infty$, we can neglect the subleading entropic term.
Eq.~(\ref{eq:PnsRatioMultInt}) becomes~:
\begin{align}
	\label{eq:PnsRatioPathInt}
	&\P(s) \simeq \e^{- N\,C_\kappa} \\ 
	&\frac{ \displaystyle
		 \int \dd c \int \D \rho \:
			\e^{-\frac{\beta N^2}{2} \mathscr{E}[\rho] }
		\delta \left(
			 \int_c \rho - \kappa
		\right)
		\delta \left(
			 \int^c \rho - (1-\kappa)
		\right)
		\delta \left(
			s - \int_c \rho(x) f(x) \: \dd x
		\right)
	}
	{ \displaystyle
		\int \dd c \int \D \rho \:
			\e^{-\frac{\beta N^2}{2} \mathscr{E}[\rho] }
		\delta \left(
			 \int_c \rho - \kappa
		\right)
		\delta \left(
			 \int^c \rho - (1-\kappa)
		\right)
	}
	\nonumber
		\:,
\end{align}
where the factor $\e^{- N\,C_\kappa}$, with $C_\kappa=-\kappa\ln\kappa-(1-\kappa)\ln(1-\kappa)$, arises from the sum over occupations in the denominator of \eqref{eq:PnsRatioMultInt} (cf. discussion in Section~\ref{sec:TypFluct}).
This additionnal factor would be absent if one would consider the distribution of the TLS restricted to the largest eigenvalue, like in Ref.~\cite{GraMajTex17}~: 
\eqref{eq:PnsRatioPathInt} without $\e^{- N\,C_\kappa}$ would describe the full range and would be clearly normalized.
Here, this form only describes the \textit{tail} of the distribution of the TLS without restriction on the ordering of the eigenvalues, when $s<s_0(\kappa)$, hence the expression has no reason to be normalised.
These integrals can be estimated using a saddle point method. The saddle point will give $\rho$ and $c$ as functions of $\kappa$ and $s$. 
The numerator is dominated by the density $\rho^\star(x;\kappa,s)$ which minimizes the energy $\mathscr{E}[\rho]$ under the constraints imposed by the Dirac $\delta$-functions. 
Similarly, the denominator is dominated by a density $\rhoMP(x)$.
The minimization takes the form similar to (\ref{eq:FCoulomb},\ref{eq:MinFapp}), with additional constraints, which leads to two coupled integral equations, instead of one \eqref{eq:MinFapp}.
The precise equations can be found in Ref.~\cite{GraMajTex17}, where this procedure was carried out explicitly.
Here, we will simply jump to the solution of the coupled integral equations (see below Eq.~\eqref{eq:rhoSpl}).
%use the results given in this paper.\\
Finally, we obtain the distribution %$\P$ is given by
\begin{equation}
  \boxed{
	\P(s) \underset{N \to \infty}{\sim} 
	    \e^{- N\,C_\kappa }
		\exp \left\lbrace - \frac{\beta N^2}{2} \Phi_-(\kappa;s)  \right\rbrace
	\hspace{0.5cm}
	\text{for}
	\hspace{0.5cm}
	 s < s_0(\kappa)
	}
	%\:,
	\label{eq:PnsLDFL}
\end{equation}
where we have introduced the large deviation function
\begin{equation}
	\Phi_-(\kappa;s) = \mathscr{E}[\rho^\star(x;\kappa,s)] - \mathscr{E}[\rhoMP(x)]
	\:.
	\label{eq:LDFLenegy}
\end{equation}
In \eqref{eq:PnsLDFL}, the scaling as $N^2$ is a manifestation of energy of the Coulomb gas and reflects the long range nature of the interaction between the $N$ particles.

%%%%%%%%%%%%%%%%%%%%%%%%%%%%%%%%%%%%%%%%%%%%%%%%%%%%%%%%%%%%%%%%%%%%%%%%%%%%%%%%%%%%%%%%%%%%%%%
\subsection{Optimal density}

\begin{figure}[!ht]
	\centering
	\includegraphics[width=0.6\textwidth]{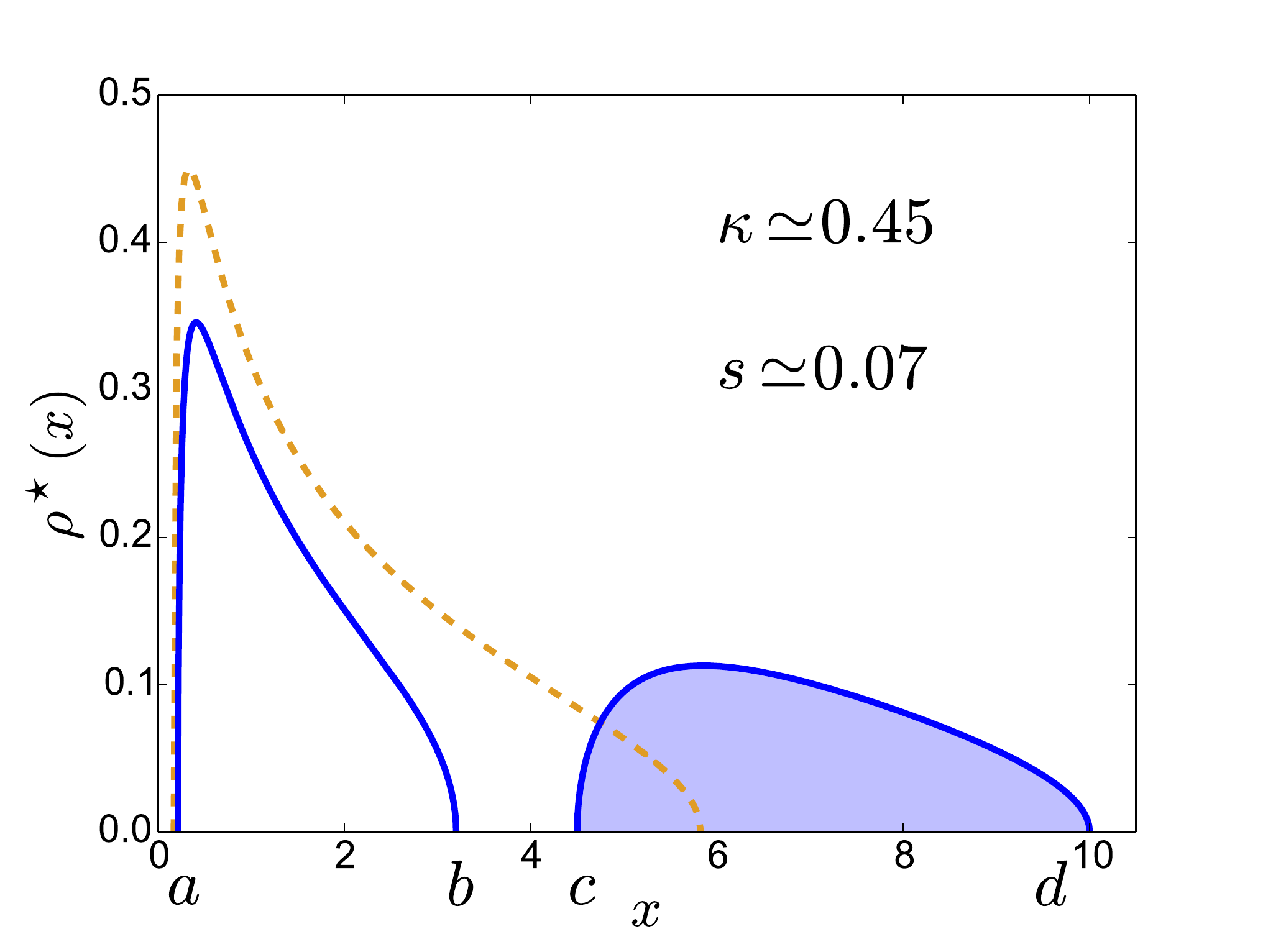}
	\caption{\it 
	 Optimal density $\rho^\star$ (solid line), for $f(x)=1/x$, compared to the density $\rhoMP$ (dashed).
	 The shaded area correspond the fraction $\kappa$ of the largest eigenvalues.
	 The support of the density $\rho^\star$ is $[a,b]\cup[c,d]$.
	  }
	\label{fig:RhoSplit}
\end{figure}

The density $\rho^\star$ is given explicitly in the general case in the appendix of Ref.~\cite{GraMajTex17}. For any monotonic function $f$, it reads~:
\begin{align}
	\label{eq:rhoSpl}
	\rho^\star(x;\kappa,s) = & \frac{1}{2\pi} \sqrt{\frac{(b-x)(d-x)}{(x-a)(c-x)}} 
	\\
	& \times \left\lbrace
		1 - \frac{1}{x} \sqrt{\frac{ac}{bd}} + \mu_1
		\dashint_c^d \frac{\dd t}{\pi} \frac{f'(t)}{t-x} \sqrt{\frac{(t-a)(t-c)}{(t-b)(d-t)}}
	\right\rbrace
	\:,
	\nonumber
\end{align}
where the principal value is needed only if $x \in [c,d]$. This density is supported on two disjoint intervals $[a,b]$ and $[c,d]$ (see Fig.~\ref{fig:RhoSplit}), where $c$ is the boundary introduced above in Eqs.~(\ref{eq:ConstrScoulomb},\ref{eq:ConstrKcoulomb}). The boundaries of these intervals and the parameter $\mu_1$ are fixed by the conditions that the density vanishes at these points~:
\begin{align}
	1 -  \sqrt{\frac{a}{bcd}} + \mu_1 \int_c^d \frac{\dd t}{\pi} f'(t) \sqrt{\frac{t-a}{(t-b)(t-c)(d-t)}} &=0
	\:,
	\label{eq:RhoCond1}
\\
	3 + \frac{a+c-b-d}{2} - \sqrt{\frac{ac}{bd}}
	- \mu_1 \int_c^d \frac{\dd t}{\pi} f'(t) \sqrt{\frac{(t-a)(t-c)}{(t-b)(d-t)}} &= 0
	\: ,
	\label{eq:RhoCond2}
\\
	1 - \sqrt{\frac{c}{a b d}} + \mu_1 \int_c^d \frac{\dd t}{\pi} f'(t) \sqrt{\frac{t-c}{(t-a)(t-b)(d-t)}} &= 0
	\:,
	\label{eq:RhoCond3}
\end{align}
along with the constraints
\begin{equation}
	\int_c^d \rho^\star(x;\kappa,s) \dd x = \kappa \:,
	\hspace{1cm}
	\int_c^d \rho^\star(x;\kappa,s) f(x) \dd x = s\:.
	\label{eq:RhoCond4}
\end{equation}
The parameter $\mu_1$ is a Lagrange multiplier introduced to handle the constraint on $s$. We denote $\mu_1^\star(\kappa;s)$ the solution of (\ref{eq:RhoCond1},\ref{eq:RhoCond2},\ref{eq:RhoCond3},\ref{eq:RhoCond4}). Its knowledge allows to compute easily the energy thanks to the thermodynamic identity \cite{GraTex15,CunFacViv16,GraTex16b}~:
\begin{equation}
	\deriv{\mathscr{E}[\rho^{\star}(x;\kappa,s)]}{s} = - \mu_1^\star(\kappa;s)
	\:.
	\label{eq:ThermoIdCoulomb}
\end{equation}

The density $\rhoMP$ which dominates the denominator of (\ref{eq:PnsRatioPathInt}) is the well known Mar\v{c}enko-Pastur distribution \cite{MP67}~:
\begin{equation}
    \label{eq:MarcenkoPastur}
	\rhoMP(x) = \frac{1}{2\pi x} \sqrt{(x-x_-)(x_+ - x)}
	\:,
	\hspace{1cm}
	x_{\pm} = 3 \pm 2 \sqrt{2}
	\:.
\end{equation}
This density can be obtained from (\ref{eq:rhoSpl}) by taking the limit $\mu_1 \to 0+$, which corresponds to $c-b \to 0$.

This derivation is valid when only the largest eigenvalues contribute to $s$. This is the case as long as the gap between the two intervals of the support of $\rho^\star$ is non-zero.
The limit of validity is therefore given by $b=c$, which corresponds to $\mu_1 = 0$. The distribution is then $\rhoMP$, thus Eq.~(\ref{eq:RhoCond4}) gives the value of $s$~:
\begin{equation}
	s_0(\kappa) = \int_{c_0}^{x_+} \rhoMP(x) f(x) \dd x
	\:,
\end{equation}
where $c_0$ is fixed by
\begin{equation}
	\kappa = \int_{c_0}^{x_+} \rhoMP(x) \dd x
	\:.
	\label{eq:constKc0}
\end{equation}
Solving the second equation for $c_0$ and plugging the result into the first equation gives the maximal value $s_0(\kappa)$ allowed for $s$. This defines a line $s = s_0(\kappa)$ in the $(\kappa,s)$ plane which delimits the region where the assumption that only the largest eigenvalues contribute to $s$ is true. This is the lower solid line in Fig.~\ref{fig:PhDiag}. Exactly on this line, the density of eigenvalues is the Mar\v{c}enko-Pastur distribution $\rhoMP$.

%%%%%%%%%%%%%%%%%%%%%%%%%%%%%%%%%%%%%%%%%%%%%%%%%%%%%%%%%%%%%%%%%%%%%%%%%%%%%%%%%%%%%%%%%%%%%%%
\subsection{Tail $s \to 0$}

For $f(x)=1/x$, the limit $s \to 0$, corresponds to push the fraction $\kappa$ of the rightmost eigenvalues towards infinity. This means to let $d > c \to \infty$ in the previous equations. Expanding Eqs.~(\ref{eq:RhoCond1},\ref{eq:RhoCond2},\ref{eq:RhoCond3},\ref{eq:RhoCond4}) in this limit yields~:
\begin{align}
	a &= 3 - 2 \sqrt{(1-\kappa)(2-\kappa)} - 2 \kappa + \O(\sqrt{s}) \:,\\
	b &= 3 + 2 \sqrt{(1-\kappa)(2-\kappa)} - 2 \kappa + \O(\sqrt{s}) \:,\\
	c &= \frac{\kappa}{s} \left(1 - \sqrt{2s} + \frac{5s}{4} + \O(s^{3/2}) \right) \:,\\
	d &= \frac{\kappa}{s} \left(1 + \sqrt{2s} + \frac{5s}{4} + \O(s^{3/2}) \right) \:,\\
	\mu_1^\star &= \frac{\kappa^2}{s^2} - \frac{3\kappa(2-\kappa)}{2s} + \O(s^{-1/2}) \:.
\end{align}
Using the thermodynamic identity (\ref{eq:ThermoIdCoulomb}), a simple integration of this last relation gives the behaviour of the energy, therefore of the large deviation function, for $s \to 0$~:
\begin{equation}
	\Phi_-(\kappa;s) = \mathscr{E}[\rho^\star(x;\kappa,s)] - \mathscr{E}[\rhoMP(x)]
		= \frac{\kappa^2}{s} + \frac{3 \kappa(2-\kappa)}{2} \ln s + \O(1) 
	\:.
	\label{eq:LDFsZero}
\end{equation}
This expression gives the left tail of the distribution~:
\begin{equation}
	\P(s) \underset{s \to 0}{\sim} 
	s^{- 3 \beta \kappa (2-\kappa) N^2/4} \: \e^{-\beta (N \kappa)^2/(2s)}
	\:.
\end{equation}
Note that we recover the tail computed in Ref.~\cite{TexMaj13} simply by setting $\kappa=1$~:
\begin{equation}
	P_{N,\kappa=1}(s) \underset{s \to 0}{\sim} s^{-3 \beta N^2/4} \: \e^{- \beta N^2/(2s)}
	\:. 
\end{equation}

Interestingly, the leading term in (\ref{eq:LDFsZero}) can be obtained easily by a heuristic argument. For small $s$, the energy is dominated by the potential energy of the eigenvalues pushed to infinity. The typical value of these eigenvalues is given by $s \sim \kappa/x_\mathrm{typ}$, corresponding to $x_\mathrm{typ} \sim \kappa/s$. The energy is estimated as $\mathscr{E}[\rho^\star(x;\kappa,s)] \sim \int_c \rho^\star V \sim \kappa V(x_\mathrm{typ}) \sim \kappa^2/s$. \textsc{Qed}.

\subsection{Limit $s\to s_0(\kappa)$}

In the limit $s\to s_0(\kappa)$, the two bulks of the density $\rho^\star$ merge. Hence, it corresponds to $c-b \to 0$.
The behaviour of $\Phi_-(\kappa;s)$ for $s$ close to $s_0(\kappa)$ is obtained by expanding 
Eqs.~(\ref{eq:rhoSpl},...,\ref{eq:RhoCond4}) 
%Eqs.~(\ref{eq:rhoSpl},\ref{eq:RhoCond1},\ref{eq:RhoCond2},\ref{eq:RhoCond3},\ref{eq:RhoCond4}) 
in this limit.
A straightforward but cumbersome computation gives, for any monotonic function $f$~:
\begin{equation}
	\mu_1^\star(\kappa;s) \simeq -2 \omega_\kappa\, (s-s_0(\kappa)) \:,
\end{equation}
where
\begin{equation}
	\omega_\kappa^{-1} =
		\int_{c_0}^{x_+} \frac{\dd x}{\pi} \frac{f(x)-f(c_0)}{\sqrt{(x-x_-)(x_+-x)}}
		\dashint_{c_0}^{x_+} \frac{\dd t}{\pi} f'(t) \frac{\sqrt{(t-x_-)(x_+-t)}}{x-t} \:,
	\label{eq:coefQuadBehav}
\end{equation}
where $c_0$ is fixed by (\ref{eq:constKc0}).
The large deviation function is deduced from the thermodynamic identity (\ref{eq:ThermoIdCoulomb})~:
\begin{equation}
	\Phi_-(\kappa;s) \simeq \omega_\kappa\, (s-s_0(\kappa))^2 \:.
	\label{eq:LDFphi0QuadMin}
\end{equation}

%
%The universal scenario presented in \cite{GraMajTex17} has shown that the large deviation function presents the quadratic behaviour when the two bulks merge~:
%\begin{equation}
%  \Phi_-(s) \underset{s\to s_0^-}{\simeq} \omega\, (s-s_0(\kappa))^2
%\end{equation}
%The coefficient $\omega$ is here difficult to obtain as the expansion of the equations fixing the different parameters for $s\to s_0^-$ is tedious.

\begin{figure}[!ht]
\centering
\includegraphics[width=0.6\textwidth]{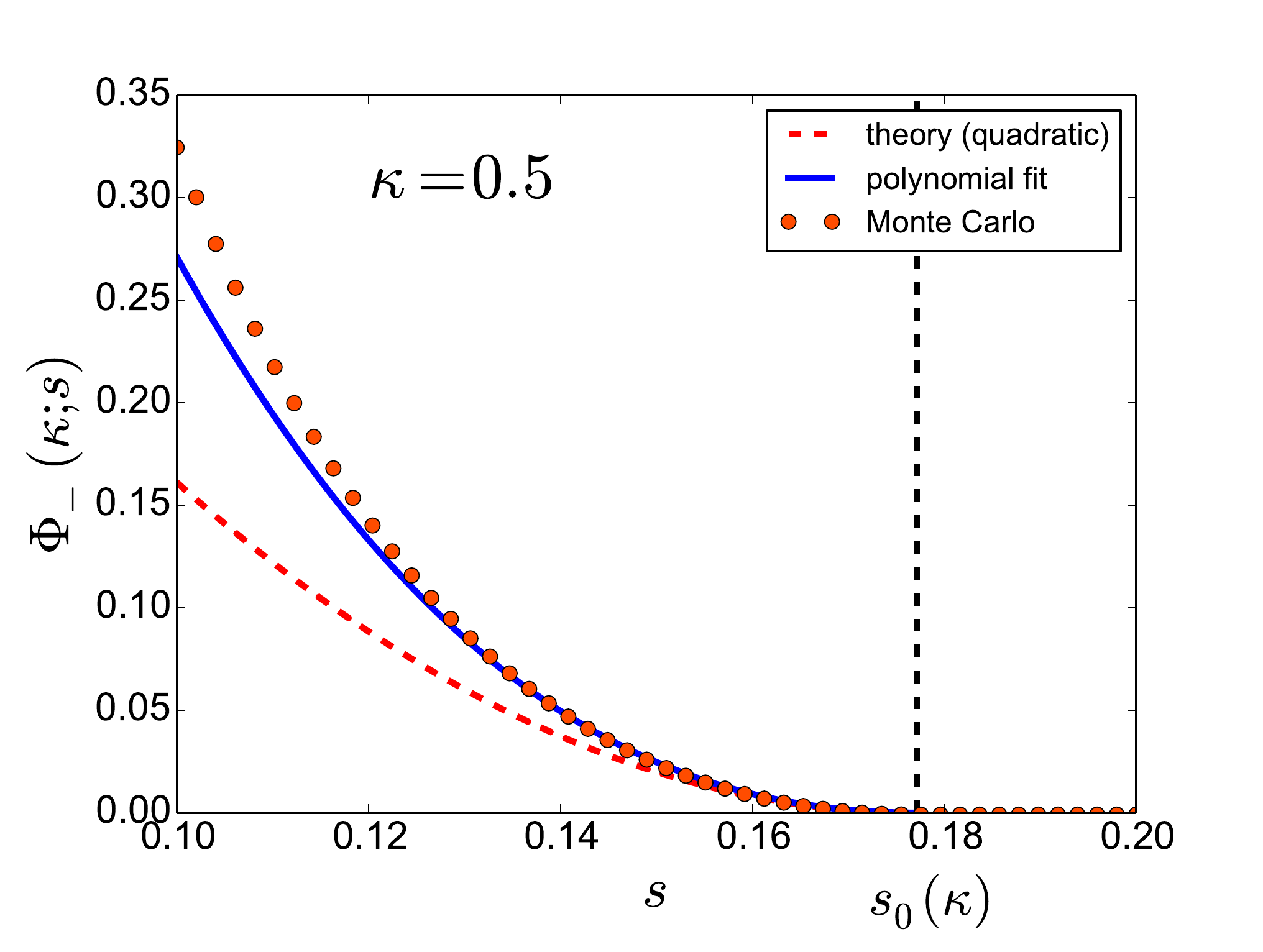}
\caption{\it Large deviation function $\Phi_-(\kappa;s) = \mathscr{E}[\rho^\star(x;\kappa,s)] - \mathscr{E}[\rho_0^\star(x)]$ (energy of the Coulomb gas) determined by a Monte Carlo simulation for $\kappa=0.5$ and $N=1500$ (dots). The results are compared to the quadratic behavior (\ref{eq:LDFphi0QuadMin}) (dashed line) and a polynomial fit of degree 3, Eq.~\eqref{eq:FitPoly3}, (solid line).
The dashed vertical line is $s_0(\kappa)$.
}
\label{fig:MC}
\end{figure}

We have performed a numerical simulation in order to check the quadratic behaviour and the value of the coefficient $\omega_\kappa$, in the case $f(x) = 1/x$.
The energy $\mathscr{E}[\rho^\star(x;\kappa,s)]$ is computed using a Monte Carlo method, see Ref.~\cite{GraMajTex17} for details on the procedure.
Fitting the numerics must be performed with caution~:
for $\kappa=0.5$, Eq.~(\ref{eq:coefQuadBehav}) gives $\omega_{1/2} \simeq 27.1$,
however a fit of the numerical data with a purely quadratic behaviour leads to $\omega_{1/2}^\mathrm{num} \simeq 32.4$.
This apparent discrepancy is explained by the fact that \eqref{eq:LDFphi0QuadMin} is only the leading term of an expansion near $s_0(\kappa)$~:
\begin{equation}
  \label{eq:FitPoly3}
	\Phi_-(\kappa;s) \simeq \omega_\kappa\, (s-s_0(\kappa))^2 + \omega_{\kappa,3} (s-s_0(\kappa))^3 + \cdots 
	\:,
\end{equation}
where the higher order term cannot be neglected in practice as it involves a large coefficient.
Indeed, fitting the numerics with both the quadratic and the cubic terms, we find 
$\omega_{1/2}^\mathrm{num} \simeq 27.0 $ and $\omega_{1/2,3}^\mathrm{num}  \simeq -239$, now in excellent agreement with the prediction
(see fit in Fig.~\ref{fig:MC}).
We have checked that the accuracy on $\omega_{1/2}^\mathrm{num}$ extracted from the fit is improved by increasing the order of the polynomial used to fit the data.

Interestingly, the simulation shows that the energy is frozen $\mathscr{E}[\rho^\star(x;\kappa,s)] = \mathscr{E}[\rho_0^\star(x)]$ for $s > s_0(\kappa)$. This will be explained in the next section.

%%%%%%%%%%%%%%%%%%%%%%%%%%%%%%%%%%%%%%%%%%%%%%%%%%%%%%%%%%%%%%%%%%%%%%%%%%%%%%%%%%%%%%%%%%
%%%%%%%%%%%%%%%%%%%%%%%%%%%%%%%%%%%%%%%%%%%%%%%%%%%%%%%%%%%%%%%%%%%%%%%%%%%%%%%%%%%%%%%%%%

\section{Phase II ($s_0(\kappa) < s < s_1(\kappa)$)~: frozen Coulomb gas and fictitious fermions}
\label{sec:TypFluct}

\begin{figure}[!ht]
	\centering
	\includegraphics[width=0.45\textwidth]{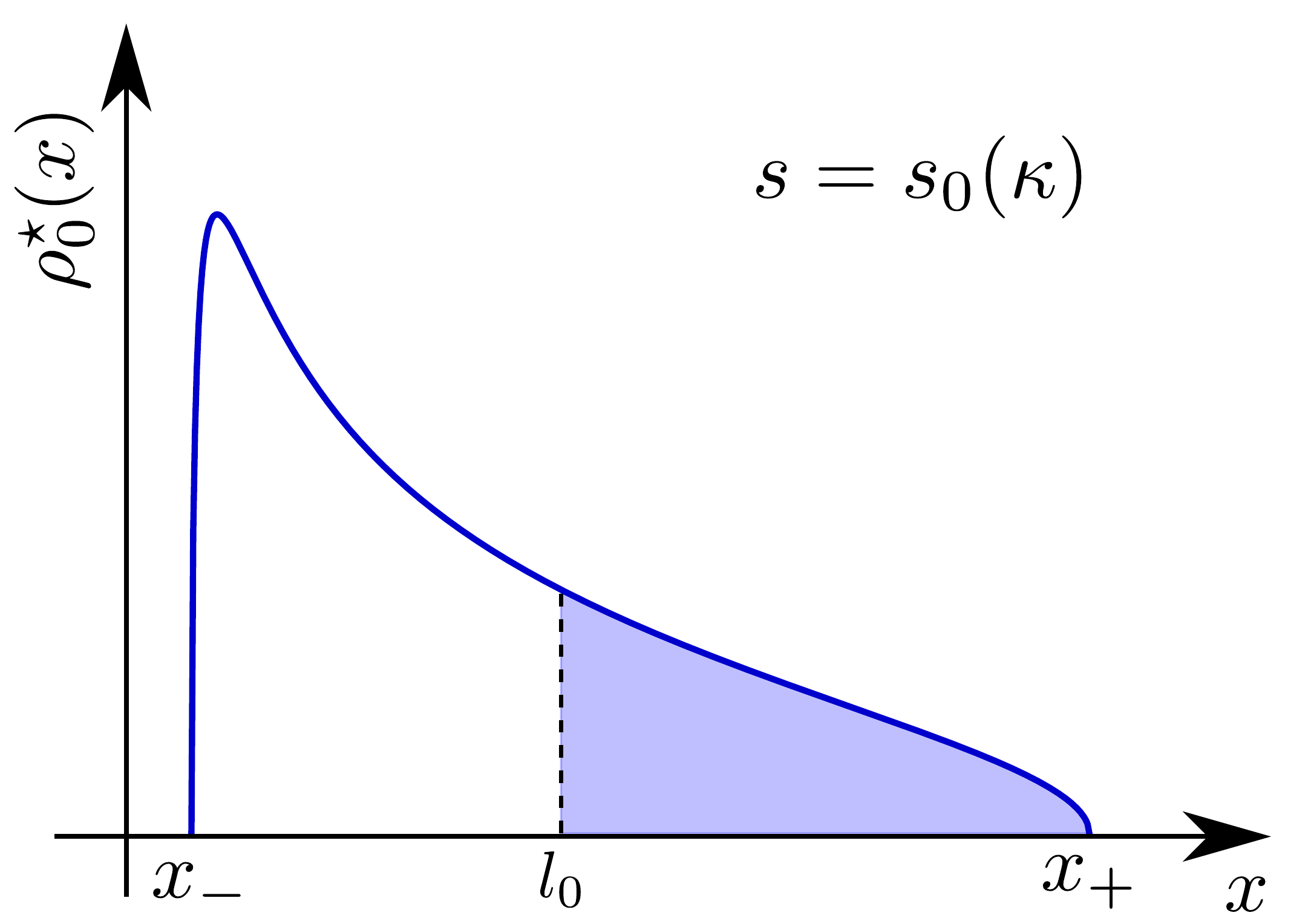}
	\includegraphics[width=0.45\textwidth]{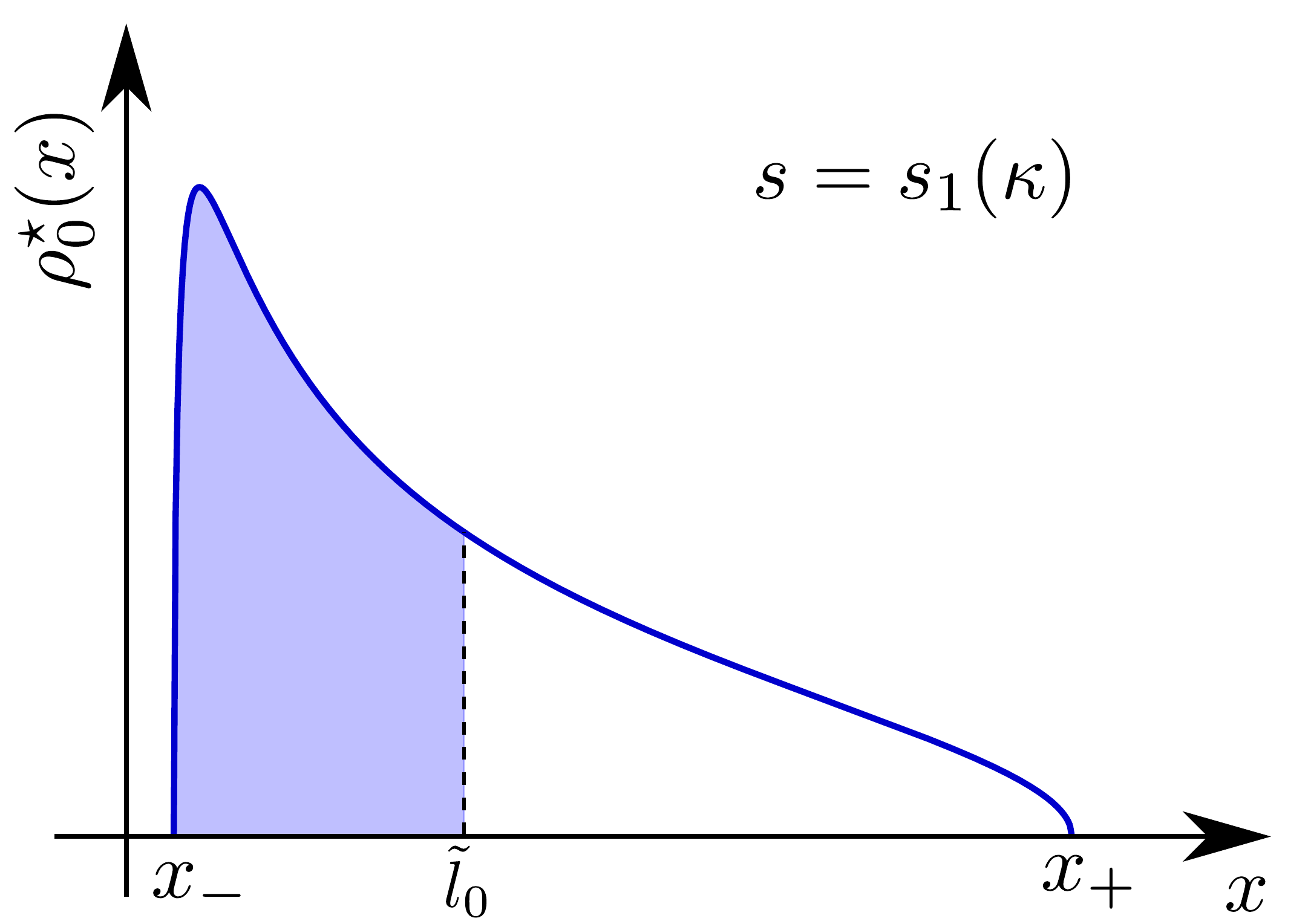}
	\caption{\it 
	 Mar\v{c}enko-Pastur distribution $\rho_0^\star$, with either the largest eigenvalues selected (left), corresponding to $s=s_0(\kappa)$, or the smallest eigenvalues (right), corresponding to $s=s_1(\kappa)$.
	  }
	\label{fig:MPminmax}
\end{figure}

In the previous section, we have seen that 
when $s\in]0,s_0(\kappa)]$, the density $\rho^\star$ changes with $s$ while the occupation numbers $\lbrace \m_i \rbrace$ remain fixed. 
%varying $s < s_0(\kappa)$ changes the density $\rho^\star$ while keeping the occupation numbers $\lbrace \m_i \rbrace$ fixed. 
When $s$ reaches $s_0(\kappa)$, the density is given by the Mar\v{c}enko-Pastur distribution $\rhoMP$, which minimizes the energy $\mathscr{E}[\rho]$.
We now enter the domain $s > s_0(\kappa)$ where it is possible to vary $s$ by changing the occupation numbers, while keeping the density fixed.
The range of accessible values of $s$ is given by two extreme cases~: selecting either the largest eigenvalues ($s = s_0(\kappa)$) or the smallest ones (Fig.~\ref{fig:MPminmax}, right). This latter case corresponds to the value $s = s_1(\kappa)$ given by~:
\begin{equation}
	s_1(\kappa) = \int_{x_-}^{c_1} \rhoMP(x)f(x) \dd x
	\:,
\end{equation}
where $c_1$ is fixed by
\begin{equation}
	\kappa = \int_{x_-}^{c_1} \rhoMP(x) \dd x
	\:;
\end{equation}
we recall that $x_\pm=3\pm2\sqrt{2}$ are the boundaries of the support of the Mar\v{c}enko-Pastur distribution introduced above, Eq.~\eqref{eq:MarcenkoPastur}.
This defines another line in the $(\kappa,s)$ plane. It is the upper solid line shown in Fig.~\ref{fig:PhDiag}.
The two lines are simply related via the relation~:
\begin{equation}
	s_1(\kappa) = \Sc - s_0(1-\kappa) \:,
	\hspace{0.5cm}
	\text{where}
	\hspace{0.5cm}
	\Sc = \int_{x_-}^{x_+} \rhoMP(x) f(x) \dd x
	\:.
\end{equation}
In the Laguerre ensemble considered here, with $f(x) = 1/x$, we simply have $\Sc = 1$.

The goal of this section is to describe what occurs for $s$ between $s_0(\kappa)$ and $s_1(\kappa)$, and compute the distribution in this domain.
Note that the procedure exposed in this section is valid for a monotonic function~$f$.

\subsection{Distribution of $s$ between $s_0(\kappa)$ and $s_1(\kappa)$~: an entropic contribution}
\label{subsec:discDistEntropy}

In order to determine the distribution $\P(s)$ on the interval $[s_0(\kappa),s_1(\kappa)]$, let us go back to the multiple integrals in Eq.~(\ref{eq:PnsRatioMultInt}). For large $N$, these integrals are dominated by the set of eigenvalues $\lbrace x_i^\star \rbrace$ which minimizes the energy  $\mathcal{E}_\mathrm{gas}[\lbrace x_i \rbrace]$ given by Eq.~(\ref{eq:Ediscr}), i.e.
\begin{equation}
	\left. \derivp{\mathcal{E}_\mathrm{gas}[\lbrace x_i \rbrace]}{x_i} 
	\right|_{ \lbrace x_i^\star \rbrace } = 0
	\:,
	\hspace{0.5cm}
	\forall i \in \lbrace 1, \cdots , N \rbrace
	\:.
	\label{eq:MinEdiscr}
\end{equation}
Without changing this set of eigenvalues, it is possible to construct $\binom{N}{\Nt}$ different values, given by 
\begin{equation}
	\frac{1}{N} \sum_{i=1}^{N} \m_i f(x_{i}^\star)
	\:.
\end{equation}
However, several configurations may give a similar value of $s$. This shows that the distribution of $s$ comes from an entropic contribution $\Entropy(\kappa;s)$~: the same ``macroscopic'' value $s$ may be associated to many ``microscopic'' configurations $\lbrace \m_i \rbrace$.
We stress that this should not be confused with the entropy $\mathscr{S}[\rho]$ defined by Eq.~(\ref{eq:Srho}). Indeed, this latter measures the loss of information when describing the set of eigenvalues $\lbrace x_i \rbrace$ by a density $\rho(x)$.
The entropy $\Entropy(\kappa;s)$ discussed here is hidden in the sums over the configurations $\lbrace \m_i \rbrace$ in (\ref{eq:PnsRatioMultInt}).

%This entropy is not related to the entropic term introduced in the previous section  when we substituted path integrals to the multiple integrals (\ref{eq:MeasurePathInt}). The entropy  defined by Eq.~(\ref{eq:Srho}) is associated to the arrangement of the $N$ eigenvalues described by the density $\rho$. Whereas here we consider the entropy $\Entropy(\kappa;s)$ associated to the number of configurations $\lbrace \m_i \rbrace$ corresponding to the value $s$. This entropy is not present in (\ref{eq:PnsRatioPathInt}), but is hidden in the sums over the configurations $\lbrace \m_i \rbrace$ in (\ref{eq:PnsRatioMultInt}).

It is more convenient to first compute the moment generating function, which is the Laplace transform of $\P(s)$, using standard tools from statistical physics.
In a second step, we will come back to the distribution.

%
%\begin{figure}[!ht]
%	\centering
%	\includegraphics[width=0.45\textwidth]{MP_max}
%	\includegraphics[width=0.45\textwidth]{MP_min}
%	\caption{\it 
%	 ...
%	  }
%	\label{fig:MPminmax}
%\end{figure}
%

%%%%%%%%%%%%%%%%%%%%%%%%%%%%%%%%%%%%%%%%%%%%%%%%%%%%%%%%%%%%%%%%%%%%%%%%%%%%%%%%%%%%%%%%%%%%%%%
\subsection{Moment generating function~: mapping to free fermions}
\label{subsec:MomentsGenFct}

The moment generating function is given by the Laplace transform of $\P$~:
\begin{align}
	\G(\t) &= \int e^{- \t N s} \P(s) \dd s = \mean{ \exp \left( - \t \sum_i \m_i f(x_i^\star) \right) }	\:, \nonumber
	\\
	&= \frac{\Nt ! (N-\Nt)!}{N!} \sum_{ \lbrace \m_i \rbrace } \prod_{i=1}^N \e^{- \t \m_i f(x_i^\star)}
	\,
	\ConstN
	\:,
	\label{eq:defGenFun}
\end{align}
%\begin{equation}
%	\G(\t) = \mean{ \exp \left( - \t \sum_i \m_i f(x_i^\star) \right) }
%		= \frac{\Nt ! (N-\Nt)!}{N!} \sum_{ \lbrace \m_i \rbrace } \prod_{i=1}^N \e^{- \t \m_i f(x_i^\star)}
%		\:,
%	\label{eq:defGenFun}
%\end{equation}
where the sum runs over the $\binom{N}{\Nt}$ configurations $\lbrace \m_i \rbrace$.
The problem of choosing $\Nt$ eigenvalues among $N$, with the constraint that $s$ is fixed, is equivalent to placing $\Nt$ particles on $N$ ``energy levels'' $\varepsilon_i = f(x_i^\star)$, with the constraint that the total ``energy'' is fixed.
Since an eigenvalue can be chosen only once, an ``energy level'' can host only one particle. Therefore these particles behave like $\Nt$ fermions.
The parameters $\kappa$ and $s$ are thus related to two fundamental properties of the fermion gas~:
$$
	\kappa \leftrightarrow \text{number of fermions} 
	\hspace{1cm}
	s \leftrightarrow \text{energy of the fermions}
$$
The set $\lbrace \m_i \rbrace$ coincides with the occupation numbers of the energy levels~: 
$\m_i = 1$ if level $\varepsilon_i$ is occupied and $\m_i = 0$ otherwise.
%Since there are $\Nt = \kappa N$ particles, $\sum_i \m_i = \Nt$.

With this interpretation, (\ref{eq:defGenFun}) is simply related to the canonical partition function
\begin{equation}
	\Z(\t) = \sum_{ \lbrace \m_i \rbrace } \e^{- \t E_\mathrm{ferm}[\lbrace \m_i \rbrace] }  
		\,
		\ConstN
	\hspace{0.3cm}\mbox{where}\hspace{0.3cm}
	E_\mathrm{ferm}[\lbrace \m_i \rbrace] = \sum_{i=1}^N \m_i \varepsilon_i
	\:,
\end{equation}
via the relation
\begin{equation}
	\G(\t) = \frac{Z_{N, \Nt}(\t)}{Z_{N, \Nt}(0)}
	\:.
	\label{eq:MomParFun}
\end{equation}
The parameter $\t$ now plays the role of an inverse temperature for the fermions, not to be confused with the Dyson index $\beta \in \lbrace 1, 2, 4 \rbrace$.

It is natural to introduce the grand-canonical partition function, which can be calculated straightforwardly~:
\begin{equation}
	\Xi_N(z,\t) = \sum_{\Nt = 0}^N z^{\Nt} Z_{N,\Nt}(\t)
	= \prod_{i=1}^N	 \left( 1 + z \, \e^{-\t \varepsilon_i} \right)
	 \:,
\end{equation}
where $z$ is the fugacity.
For large $N$, the distribution of the eigenvalues $\lbrace x_i^\star \rbrace$ is the Mar\v{c}enko-Pastur law $\rhoMP$. Therefore in this limit we can write~:
\begin{align}
	\frac{1}{N} \ln \Xi_N(z,\t)
		&= \frac{1}{N} \sum_{i=1}^N \ln \left( 1 + z\, \e^{-\t \varepsilon_i} \right) \nonumber \\
		& \simeq \int_{x_-}^{x_+} \rhoMP(x) \ln \left( 1 + z\, \e^{-\t f(x)} \right) \dd x
	\:.
\end{align}
We can go back to the canonical partition function using the relation
\begin{equation}
	Z_{N,\Nt}(\t) = \frac{1}{2 \I \pi} \oint \frac{\Xi_N(z,\t)}{z^{\Nt+1}} \dd z
	\:,
	\label{eq:CanFromGCan}
\end{equation}
where the integral runs over a contour which encloses the origin once in the counter-clockwise direction.
For large $N$, this integral can be estimated using a saddle point method. In the thermodynamic limit $N \to \infty$, this corresponds to use the equivalence between the canonical and grand-canonical ensembles through a Legendre transform. The value $\zt(\kappa;s)$ of the fugacity is fixed by imposing that the mean number of fermions is $\Nt$~:
\begin{equation}
	z \left. \derivp{\ln \Xi}{z} \right|_{\zt} = p
	\hspace{0.5cm}
	\Rightarrow
	\hspace{0.5cm}
	\int_{x_-}^{x_+} \frac{\rhoMP(x)}{\exp(\t f(x))/\zt + 1} \dd x = \kappa
	\:.
	\label{eq:ConstK}
\end{equation}
Then Eq.~(\ref{eq:CanFromGCan}) yields the free energy per fermion in the thermodynamic limit~:
\begin{equation}
	\f(\kappa; \t) =- \lim_{N \to \infty}
	\frac{1}{N} \ln Z_{N,\Nt}(\t)
	= \kappa \ln \zt - \lim_{N \to \infty} \frac{1}{N} \ln \Xi_N(\zt,\t)
	\:.
\end{equation}
Note that we dropped the factor $\t$ in the usual definition of the free energy for convenience.
We get~:
\begin{equation}
	\f(\kappa; \t) = -\int_{x_-}^{x_+} \rhoMP(x) \ln \left( 1 + \zt(\kappa;\t) \, \e^{-\t f(x)} \right) \dd x
		+ \kappa \ln \zt(\kappa;\t)
	\:.
	\label{eq:FreeE}
\end{equation}
Eq.~(\ref{eq:MomParFun}) shows that the logarithm of the moment generating function, namely the cumulant generating function, is given by the free energy of the fermions~:
\begin{equation}
	\frac{1}{N} \ln \G(\t) \simeq  \f(\kappa; 0) -\f(\kappa; \t)
	\:.
	\label{eq:CumulGenFctAsymp}
\end{equation}

\subsubsection{Cumulants of the truncated linear statistics $s$, for $f(x)=1/x$}
\label{subsubsec:cumulants}

The cumulants of (\ref{eq:defTrlinstat0}) can be obtained by expanding $\ln \G$ as a power series~:
\begin{equation}
	\ln \G(\t) = \sum_{n \geq 1} \mean{s^n}_c \frac{(\t N)^n}{n!}
	\:.
	\label{eq:CumulGenFctExp}
\end{equation}
The procedure is as follows~: first determine $\zt$ as a power series in $\t$ from (\ref{eq:ConstK}). Then plug this expression into (\ref{eq:FreeE}) to get the expansion of $\f$ in powers of $\t$, and read the cumulants from the coefficients of this expansion.
In particular, for $f(x)=1/x$, the first six cumulants are given by~:
\begin{align}
	\mean{s} &= \kappa \:, \\
	\mean{s^2}_c &\simeq \frac{\kappa(1-\kappa)}{N} \:, \label{eq:Variance} \\
	\mean{s^3}_c &\simeq \frac{2 \kappa(1-\kappa)(1-2\kappa)}{N^2}
	\:, \\
	\mean{s^4}_c &\simeq \frac{2 \kappa(1-\kappa)(2-15\kappa+15\kappa^2)}{N^3} \label{eq:Cumul4} \:, \\
	\mean{s^5}_c &\simeq \frac{4\kappa(1-\kappa)(1-2\kappa)(1-42\kappa+42\kappa^2)}{N^4} \:, \\
	\mean{s^6}_c &\simeq -\frac{2\kappa(1-\kappa)(13 + 420\kappa -2940\kappa^2 +5040\kappa^3 - 2520\kappa^4)}{N^5} \label{eq:Cumul6}
	\:.
\end{align}
The first two cumulants coincide with those obtained before, Eqs.~(\ref{eq:MeanS},\ref{eq:VarS}), as it should. 

In addition, we have performed an orthogonal polynomial calculation in the case $\beta=2$ \cite{Meh04}.
This provides another check of our expressions as we have obtained that the first four cumulants given by this second method perfectly match the above expressions.

We also stress the $\kappa\leftrightarrow 1-\kappa$ symmetry of the cumulants:
under this change, the even order cumulants are unchanged, while the odd ones change sign.
Consequently, for $\kappa = 1/2$, all odd order cumulants vanish: the distribution is symmetric around its mean $s=1/2$. 
This can be understood as the symmetry between fermions and holes.

It is interesting to compare these cumulants with the cumulants of the Wigner time delay \cite{CunMezSimViv16}, corresponding to $\kappa=1$~:
\begin{align}
	\mean{s^2}_c &\simeq \frac{4}{\beta N^2} \:, \\
	\mean{s^3}_c &\simeq \frac{96}{\beta^2 N^4}
	\:, \\
	\mean{s^4}_c &\simeq \frac{5088}{\beta^3 N^6} \:, \\
	\mean{s^5}_c &\simeq \frac{437760}{\beta^4 N^8}
	\:,
\end{align}
(the variance was also found in \cite{LehSavSokSom95,BroBut97,TexMaj13}~; Ref.~\cite{MezSim13} has given the four first cumulants explicitly).

The vanishing of the cumulants (\ref{eq:Variance},...,\ref{eq:Cumul6}) for $\kappa=1$ can be straightforwardly understood from the fermionic nature of the problem~: when $\Nt = N$, the occupation numbers are all equal to one, and therefore do not fluctuate.
However, (\ref{eq:Variance},...,\ref{eq:Cumul6}) correspond to the dominant terms of expansions in powers of $1/N$. When $\kappa=1$ the leading terms vanish, and only subleading contributions remain. 
This explains why the cumulants do no present the same scaling with $N$~: 
$$
  \mean{s^n}_c \sim N^{-n+1} \mbox{ for }\kappa < 1 
  \hspace{0.5cm}\mbox{\textit{versus}}\hspace{0.5cm}
  \mean{s^n}_c \sim \beta^{-n+1} N^{-2n+2}  \mbox{ for }\kappa=1
  \:.
$$
The fluctuations are much larger when $\kappa < 1$.
This observation on the scaling of the moments has a simple interpretation~: 
the energy of the $N$ interacting particles scales as $N^2 \mathcal{E}_\mathrm{gas} \sim N^2$ due to the long range nature of the interaction, and the entropy of the $p$ fictitious fermions scales as $N \Entropy \sim N$. 
In usual studies of linear statistics, the energy dominates, and the entropy $\Entropy$ is zero because all the occupation numbers are fixed to $\m_i = 1$. Hence the fluctuations scale as $\delta s \sim 1/N$. But here, for $\kappa<1$, the energy is frozen and the additional entropy dominates. Therefore the fluctuations scale as $\delta s \sim 1/\sqrt{N}$.

This interpretation also explains why the cumulants do not depend on the Dyson index $\beta$ when $\kappa<1$~: this index is present in the  Gibbs measure $\exp[-(\beta/2)N^2\mathcal{E}_\mathrm{gas}]$, but not in the measure of entropic nature $\exp[N\Entropy]$ (as stressed in \S~\ref{subsec:discDistEntropy}, the entropy of the fermions should not be confused with the entropy of the density of the Coulomb gas).

%Note that these cumulants depend on the Dyson index $\beta$, while the ones we computed for $\kappa < 1$ do not. This was expected since this index appears \Cred{ in the energy of the Coulomb gas $\mathcal{E}_\mathrm{gas}$}, and here the {\it energy} is fixed. The distribution of $s$ takes its origin in the {\it entropy} $\Entropy$ of the random occupations $\lbrace \m_i \rbrace$. This entropy is indeed independent of $\beta$.

\subsubsection{Numerics}
\label{subsec:numerics}

\begin{figure}[!ht]
	\centering
	\includegraphics[width=0.45\textwidth]{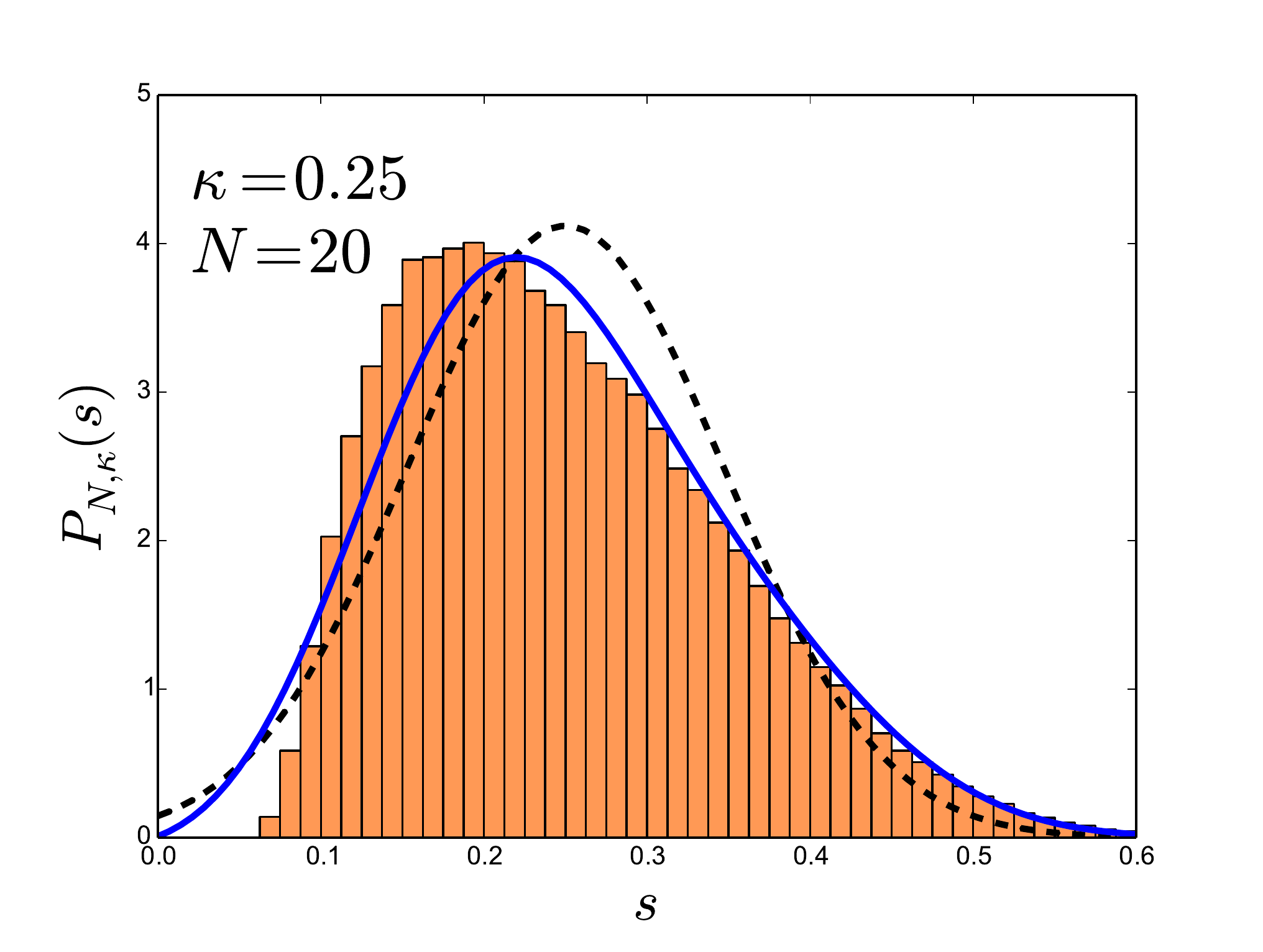}
	\hfill
	\includegraphics[width=0.45\textwidth]{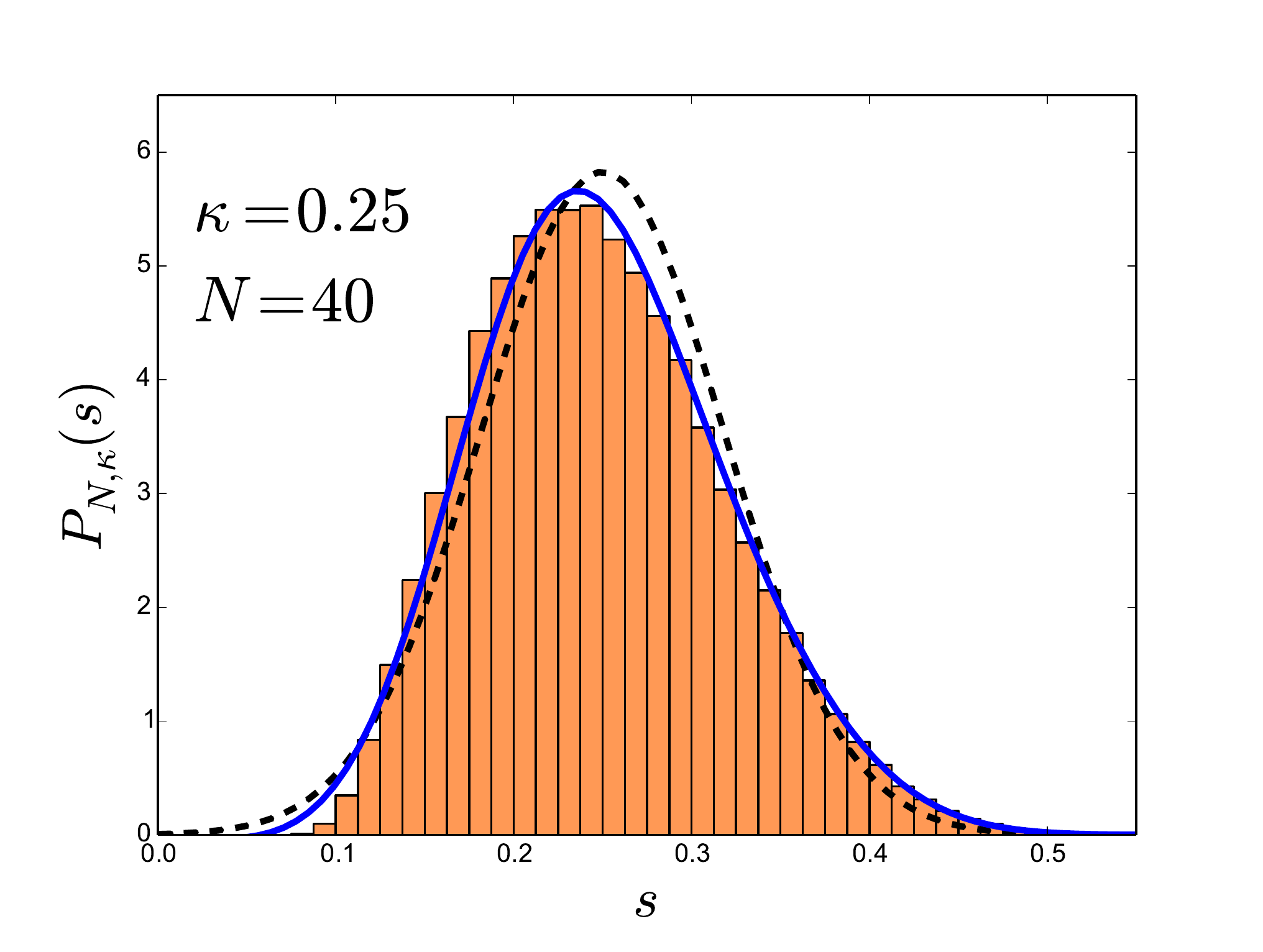}
	\caption{\it 
	Histogram of $\P(s)$, for $f(x)=1/x$,  obtained from $200\,000$ complex matrices ($\beta=2$) of various sizes, for $\kappa = 0.25$.
	The lines correspond to a reconstruction of the distribution using a Edgeworth series, using the first two (dashed) and four (solid) asymptotic forms of the cumulants. 
	  }
	\label{fig:HistNum}
\end{figure}

We have also performed numerical simulations to check our results. 
Figure \ref{fig:HistNum} shows histograms obtained 
%We have obtained histograms of the distribution $\P(s)$, \Cblu{in the case $f(x)=1/x$,} 
by diagonalizing $200\,000$ complex Hermitian matrices ($\beta = 2$) of various sizes. 
These histograms are compared to the approximate distribution reconstructed using a Edgeworth series, from the large $N$ expressions of the cumulants (\ref{eq:Variance},...,\ref{eq:Cumul4}).
On purpose, we first consider rather small matrices, of size $N=20$ (Fig.~\ref{fig:HistNum}, left)~: we can see a significant deviation, which we mostly attribute to the large $N$ approximation of the cumulants.
For still moderate size $N=40$ (Fig.~\ref{fig:HistNum}, right), we see that the agreement is already very good.
The plots of Figure~\ref{fig:HistNum} correspond to $\kappa=0.25$. 
Note that, increasing $\kappa$ with fixed $N$, we have observed that the agreement becomes less good, although cumulants are symmetric under $\kappa\leftrightarrow1-\kappa$ at leading order in $N$~; this is explained by the fact that finite $N$ corrections increase with $\kappa$, cf. Eq.~\eqref{eq:VarS}.
We have also performed the calculation up to $N=100$ where the histogram is almost indistinguishable from the simple Gaussian approximation.

%%%%%%%%%%%%%%%%%%%%%%%%%%%%%%%%%%%%%%%%%%%%%%%%%%%%%%%%%%%%%%%%%%%%%%%%%%%%%%%%%%%%%%%%%%%%%%%
\subsection{Large deviations~: fermions with positive or negative absolute temperature}
\label{subsec:ldfFermionsFD}

The distribution of the truncated linear statistics (\ref{eq:defTrlinstat0}) can be obtained in the interval $[s_0(\kappa), s_1(\kappa)]$ by Laplace inversion~:
\begin{equation}
	\P(s) = \frac{N}{2 \I \pi} \int_{\I \mathbb{R}} \e^{N \t s} \G(\t) \dd \t
	\:.
\end{equation}
For large $N$, we can estimate this integral with a saddle point method. This corresponds to performing another Legendre transform on the free energy $\f$.
In section \ref{subsec:MomentsGenFct} we already performed a Legendre transform to go from grand-canonical to canonical ensemble. Thus, here we perform a second one to reach the microcanonical ensemble where the energy (i.e. $s$) and the fermion number (i.e. $\kappa$) are fixed.

Using (\ref{eq:CumulGenFctAsymp}) for the expression of $G_{N,\Nt}$, we obtain that the saddle point $\ts(\kappa;s)$ is given by
\begin{equation}
	\left. \deriv{}{\t} \f(\kappa;\t) \right|_{\ts} - s = 0
	\:.
	\label{eq:SaddleLDF}
\end{equation}
Using the expression of $\f$, this equation reads~:
\begin{equation}
	\int_{x_-}^{x_+} \frac{\rhoMP(x)}{\exp(\ts f(x))/\zs + 1} f(x) \dd x= s \:,
	\label{eq:ConstSldf}
\end{equation}
where we denoted $\zs(\kappa;s) = \zt(\kappa; \ts(\kappa;s))$.
Then the distribution becomes~:
\begin{equation}
    \boxed{
	\P(s) 
	  \underset{N\to\infty}{\sim}
		\exp \left\lbrace
				- N \Phi_0(\kappa;s)
			\right\rbrace	
	\hspace{0.5cm}
	\text{for}
	\hspace{0.5cm}
	 s \in [s_0(\kappa),s_1(\kappa)]
	}		
	%\:,
	\label{eq:PnsPhi0}
\end{equation}
where we introduced the large deviation function $\Phi_0$, which is given by~:
\begin{equation}
	\Phi_0(\kappa;s) = \f(\kappa; \ts(\kappa;s)) - s \ts(\kappa;s) - \f(\kappa;0)
	\:.
	\label{eq:Phi0Fsbeta}
\end{equation}
Explicitly, it reads~:
\begin{align}
	\Phi_0(\kappa;s) = \kappa \ln \zs(\kappa;s) 	\nonumber
	&- \int_{x_-}^{x_+} \rhoMP(x) \ln \left( 1 + \zs(\kappa;s) \e^{-\ts(\kappa;s)f(x)} \right) \dd x
	\\
	& - s \ts(\kappa;s)
		- \kappa \ln \kappa - (1-\kappa) \ln (1-\kappa)
	\:,
	\label{eq:LdfPhi0}
\end{align}
where $\ts$ and $\zs$ are fixed by Eq.~(\ref{eq:ConstSldf}) and
\begin{equation}
	\int_{x_-}^{x_+} \frac{\rhoMP(x)}{\exp(\ts f(x))/\zs + 1} \dd x= \kappa \:.
	\label{eq:ConstKm}
\end{equation}

The observations we made earlier discussing the cumulants are also valid for the full distribution of $s$, Eqs.~(\ref{eq:PnsPhi0},\ref{eq:LdfPhi0}). For $s \in [s_0(\kappa),s_1(\kappa)]$ the distribution $\P(s)$ does not depend on the Dyson index $\beta$. Moreover, the logarithm of the probability scales as $N$ whereas it scales as $N^2$ for $s < s_0(\kappa)$, indicating a distribution of width $\sim1/\sqrt{N}$, much broader that $1/N$ for $\kappa=1$.

The physical interpretation of $\Phi_0$ is made clear by Eq.~(\ref{eq:Phi0Fsbeta})~: $s$ is the energy of the fermions, $\f$ their free energy (up to a factor $\t$) and $\ts$ is the inverse temperature. Therefore, $\Phi_0$ is a difference of entropy~:
\begin{equation}
	\Phi_0(\kappa;s) = \Entropy(\kappa; s^\star) -  \Entropy(\kappa; s)
	\:,
	\label{eq:Phi0DiffEntropy}
\end{equation}
where $\Entropy = \t s - \f$ is the entropy of the $\Nt = \kappa N$ fermions with total energy $s$. The condition $\ts(\kappa;s^\star) = 0$ determines $s^\star$. The term $\Entropy(\kappa; s^\star)$ comes from the normalization of $\P(s)$, namely the denominator in Eq.~(\ref{eq:PnsRatioMultInt}).
In particular, the number of configurations $\lbrace \m_i \rbrace$ associated to the value $s$ is exactly $\exp[N\, \Entropy(\kappa;s)]$.

The values of the inverse temperature $\ts$ and the fugacity $\zs$ are obtained in terms of $\kappa$ and $s$ by Eqs.~(\ref{eq:ConstSldf},\ref{eq:ConstKm}). This corresponds to summing over all the eigenvalues (equivalently, the energy levels), with a weight given by the Fermi-Dirac distribution
\begin{equation}
	\mean{\m_i} = \frac{1}{\exp(\t f(x_i))/z + 1} 
	\:.
	\label{eq:FermiDirac}
\end{equation}
In addition, the eigenvalues are distributed according to the optimal distribution in the absence of constraint, 
i.e. the Mar\v{c}enko-Pastur distribution $\rhoMP$.

\begin{figure}[!ht]
	\centering
	\includegraphics[width=0.45\textwidth]{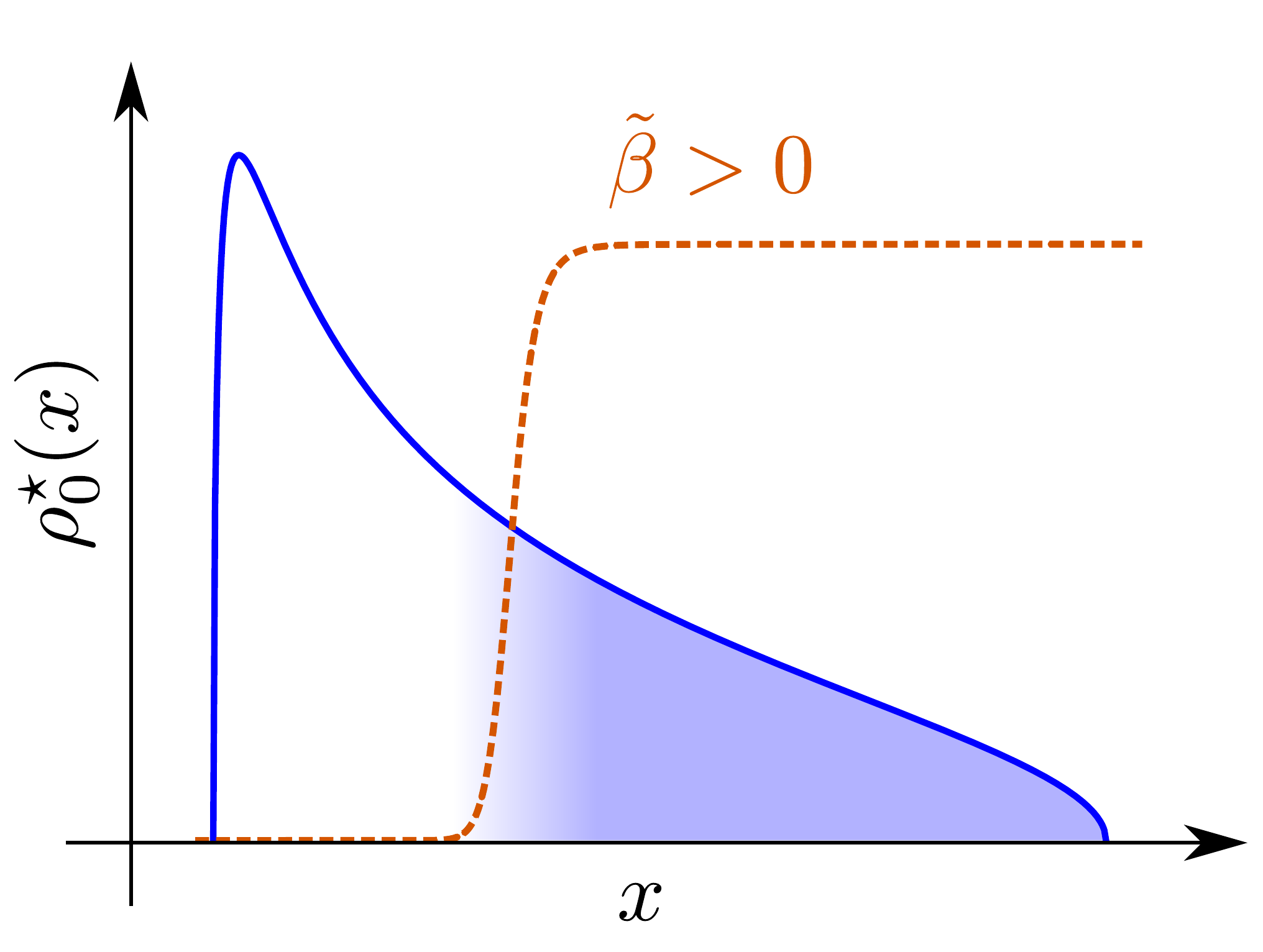}
	\includegraphics[width=0.45\textwidth]{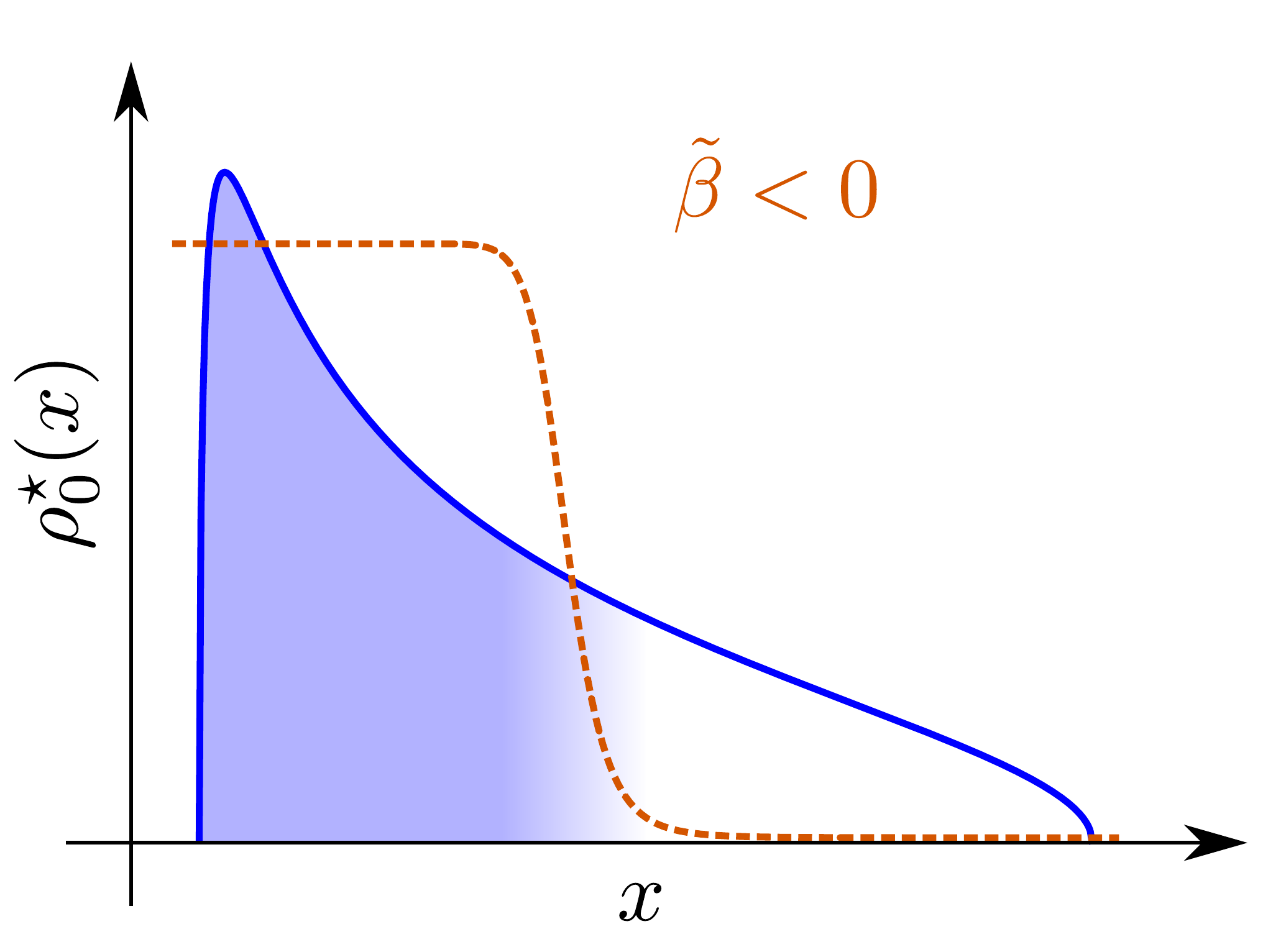}
	\caption{\it 
	 Fermi-Dirac distribution (\ref{eq:FermiDirac}) in the case $f(x)=1/x$, for $\t > 0$ (left) and $\t < 0$ (right), superimposed to the eigenvalue distribution $\rhoMP$. Shaded areas indicate the contribution of the eigenvalues to the linear statistics.
	 Note that the case $\t > 0$ corresponds to select the largest eigenvalues, hence the smallest ``energies'' $\varepsilon_n = f(x_n) = 1/x_n$, as usual.
	  }
	\label{fig:MPminmaxFD}
\end{figure}

Note that the parameter $\t$, which represents the inverse temperature of the fermions, can be either positive or negative. This quite unusual situation occurs because the spectrum $\lbrace \varepsilon_i = f(x_i^\star) \rbrace$ is bounded from below and above.
The Fermi-Dirac distribution (\ref{eq:FermiDirac}) interpolates between two step-functions~: in the limit $\t \to + \infty$, it selects only the lowest energy levels (in our case, the largest eigenvalues 
since $f$ is decreasing, as shown in Fig \ref{fig:MPminmaxFD}, left). The opposite limit $\t \to -\infty$ corresponds to select only the highest energy levels (smallest eigenvalues, see Fig \ref{fig:MPminmaxFD}, right).

The derivation we followed in this section can be generalized straightforwardly to other matrix ensembles. For example in the Gaussian ensemble, one should replace the Mar\v{c}enko-Pastur distribution $\rhoMP$ by the Wigner semicircle law.
Because the density is given by the density obtained in the absence of constraint, the scenario described here is completely universal and valid for any monotonic function~$f$ (the case of $f$ non monotonic is discussed in Section~\ref{subsec:OpenQuestions}).

% ------------------------------------------------------------------------------------------
\subsubsection{Expansion around the typical value~: infinite temperature ($\t \to 0^{\pm}$)}

The large deviation function $\Phi_0$ being obtained from a saddle point estimate, Eq.~(\ref{eq:SaddleLDF}) can be used to prove an identity similar to (\ref{eq:ThermoIdCoulomb})~:
\begin{equation}
   \boxed{
	\deriv{\Entropy(\kappa;s)}{s} =  \ts(\kappa;s)
	}
	%\:.
	\label{eq:ThermoIdPhi0}
\end{equation}
From this relation, it is clear that the maximum of $\Entropy$ (minimum of $\Phi_0$) is given by $\ts = 0$, corresponding to $s = \kappa$.
This means an infinite temperature for the fermions~: the Fermi-Dirac distribution (\ref{eq:FermiDirac}) is flat, and all the energy levels are occupied with probability $\Nt/N$.  
Expanding Eqs.~(\ref{eq:ConstSldf},\ref{eq:ConstKm}) for $\t$ near $0$ allows to compute $\ts$ near $s = \kappa $. In the case $f(x)=1/x$, this gives~:
\begin{equation}
	\ts(\kappa;s) \simeq - \frac{s-\kappa}{\kappa(1-\kappa)}
	\:.
\end{equation}
Using (\ref{eq:Phi0DiffEntropy},\ref{eq:ThermoIdPhi0}), we obtain~:
\begin{equation}
  \label{eq:Phi0closeToKappa}
	\Phi_0(\kappa;s) \simeq \frac{(s-\kappa)^2}{2 \kappa(1-\kappa)} \:,
	\hspace{0.5cm}
	\text{for } s \to \kappa \:.
\end{equation}
As we have already seen, the distribution $\P(s)$ is dominated by a Gaussian peak located at $s=\kappa$, with variance given by (\ref{eq:Variance}).
This expression of $\Phi_0$ is singular when $\kappa = 1$. But as we discussed in section \ref{subsubsec:cumulants}, this corresponds only to the leading term of an expansion in powers of $1/N$ of the variance. 
Eq.~\eqref{eq:VarS} shows that the distribtuion is regular for $\kappa\to1$~:
\begin{equation}
	-\frac{1}{N}\ln \P (s) 
	\underset{s\sim\kappa}{\simeq}
	\frac{(s-\kappa)^2}{ \displaystyle
	2 \left[ \kappa(1-\kappa)
	+ \frac{4\kappa}{\beta N} \left(1 - \beta \frac{(1-\kappa)}{4} \right)
	+ \O(N^{-2}) \right]} \:.
\end{equation}
This expression describes correctly the limit $\kappa \to 1$ for $s$ close to $\kappa$
and account for the transition between fluctuations $\sim1/\sqrt{N}$ for $\kappa<1$ and fluctuations $\sim1/N$ for $\kappa=1$ (the crossover between the two regimes obviously occurs at $1-\kappa\sim1/{\beta N}$).

% ------------------------------------------------------------------------------------------
\subsubsection{Limiting behaviours at the edges: zero temperature ($\t \to \pm \infty$)}

The limits $s \to s_0(\kappa)$ and $s \to s_1(\kappa)$ correspond to the inverse temperature $\ts \to + \infty$ and $\ts \to -\infty$, respectively.
This means zero temperature for the fermions~: the Fermi-Dirac distribution (\ref{eq:FermiDirac}) becomes a step function, selecting either the largest or the smallest eigenvalues, as shown in Fig.~\ref{fig:MPminmaxFD}.
Therefore, Eqs.~(\ref{eq:ConstSldf},\ref{eq:ConstKm}) involving integrals of this distribution can be approximated via a standard Sommerfeld expansion \cite{PatBea11,TexRou17book}~: for any function $H$ and fixed chemical potential $\mu$, one has
% \cite{AshMer76}~:
\begin{equation}
	\int \frac{H(\varepsilon) \dd \varepsilon}{\e^{\t(\varepsilon - \mu)} + 1} 
		= \int_{-\infty}^\mu H(\varepsilon) \dd \epsilon
		+ \frac{1}{\t^2} \frac{\pi^2}{6} H'(\mu) + \O(\t^{-4})
	\text{ for } \t \to +\infty
	\:,
	\label{eq:Sommerfeld}
\end{equation}
and a similar expression holds for $\t \to -\infty$.
In our case, since $\varepsilon= f(x) = 1/x$, it is convenient to introduce $\a = 1/\mu$, which delimits the domain of the spectrum of eigenvalues which is ``occupied'', as shown in Fig.~\ref{fig:MPminmax}. The fugacity is simply related to the chemical potential $\mu$ by $\zs = \e^{\ts \mu} = \e^{\ts/\a}$.

For $\ts \to + \infty$ (corresponding to $s \to s_0(\kappa)$), Eq.~(\ref{eq:ConstKm}) allows to compute $\a$ as a power series $\a = \a_0 + \a_1 \ts^{-2} + \O(\ts^{-4})$ using the Sommerfeld expansion (\ref{eq:Sommerfeld}). This gives 
\begin{equation}
	\int_{\a_0}^{x_+} \rhoMP(x) \dd x = \kappa
	\hspace{0.5cm}
	\text{and}
	\hspace{0.5cm}
	\a_1 = \frac{ \a_0^3 \pi^2 }{6} \left(
			2 + \a_0 \frac{\rhoMP{}'(\a_0)}{\rhoMP{}(\a_0)}
		\right)
	\:.
\end{equation}
Then, following the same procedure with Eq.~(\ref{eq:ConstSldf}) yields
\begin{equation}
	s = s_0(\kappa) + \frac{1}{\ts^2} \frac{\pi^2 \a_0^2 \rhoMP(\a_0)}{6} 
		+ \O(\ts^{-4})
	\:,
\end{equation}
thus
\begin{equation}
	\ts \simeq \pi \a_0 \sqrt{\frac{\rhoMP(\a_0)}{6 (s - s_{0}(\kappa)) }}
		\quad \text{as} \quad
	s \to s_{0}(\kappa)
	\:.
\end{equation}
A similar computation allows to estimate the integral in Eq.~(\ref{eq:LdfPhi0}). Finally, we obtain~:
\begin{equation} 
  \label{eq:LimitingBehaviour344}
	\Phi_0(\kappa;s) \underset{s \to s_0}{\simeq}
	     C_\kappa 
		- \pi \a_0 \sqrt{\frac{2}{3} \rhoMP(\a_0) (s - s_{0}(\kappa)) }
	\:,
\end{equation}
where $C_\kappa=-\kappa \ln \kappa - (1-\kappa) \ln (1-\kappa)>0$.
The case $\ts \to -\infty$ (corresponding to $s \to s_1(\kappa)$) can be treated in the same way. We obtain~:
\begin{equation}
  \label{eq:Phi0closeToS1}
	\Phi_0(\kappa;s) \underset{s \to s_1}{\simeq}
	  C_\kappa 
		- \pi \tilde{\a}_0 \sqrt{\frac{2}{3} \rhoMP(\tilde{\a}_0) (s_{1}(\kappa) - s) }
	\:,
\end{equation}
where $\tilde{\a}_0$ is now fixed by
\begin{equation}
	\int_{x_-}^{\tilde{\a}_0} \rhoMP(x) \dd x = \kappa \:.
\end{equation}

\begin{figure}[!ht]
	\centering
	\includegraphics[width=0.7\textwidth]{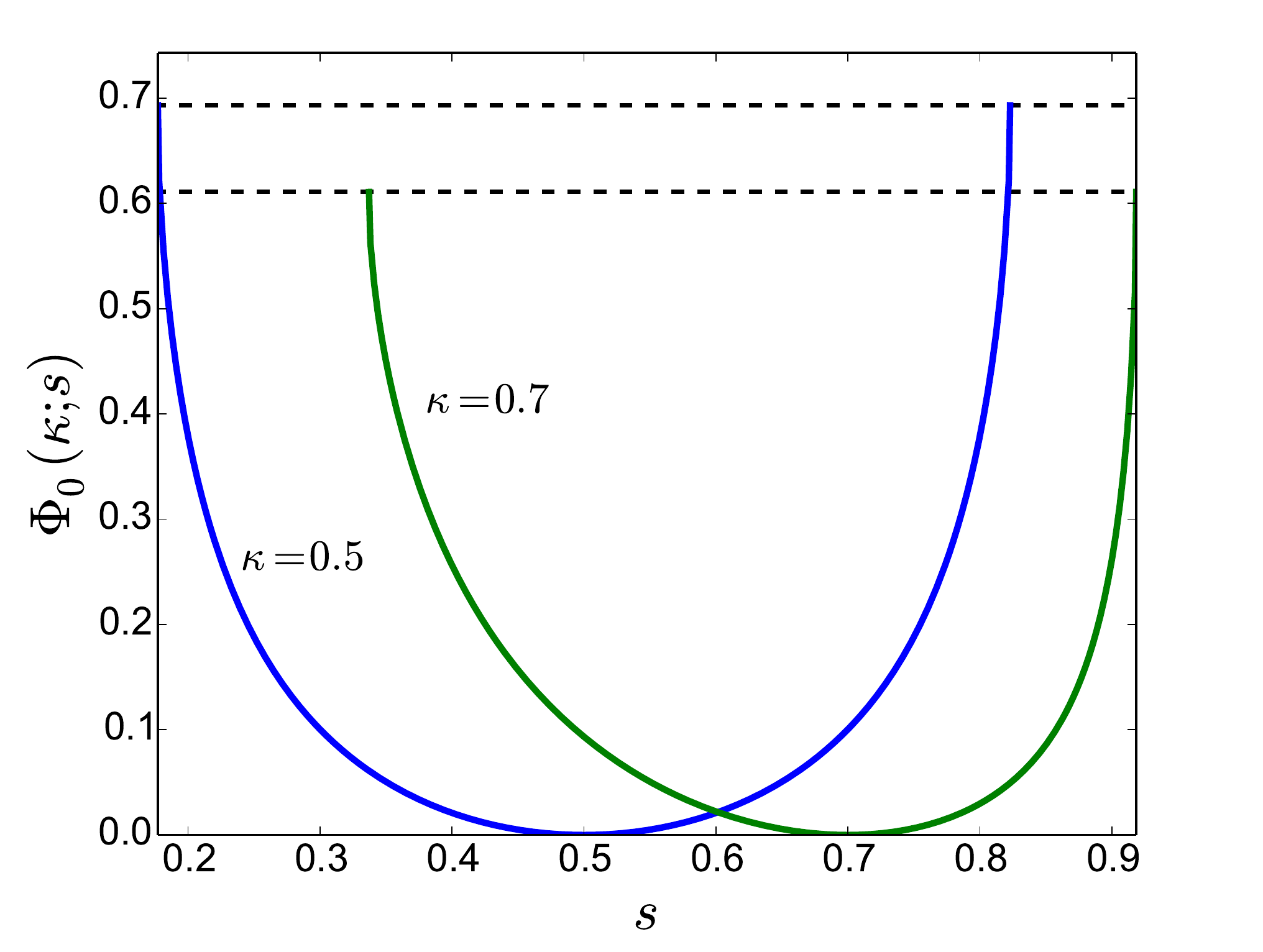}
	\caption{\it 
	 Large deviation function $\Phi_0$ (entropy of the fermions) describing in particular the typical fluctuations of the truncated linear statistics.
	 The cases $\kappa=0.5$ and $\kappa=0.7$ are both represented.
	 The dashed horizontal lines delimit the maximum $-\kappa \ln \kappa - (1-\kappa) \ln (1-\kappa)$ of these functions.
	 }	  
	\label{fig:LDFAsympt}
\end{figure}

Note that $\Phi_0$ reaches its maximal value $C_\kappa$ for $s = s_{0}(\kappa)$ and $s_1(\kappa)$. This particular value has a very simple meaning~: $\P(s)$ is the (normalized) number of configurations among the $\binom{N}{\Nt}$ which give the value $s$. For $s = s_{0}(\kappa)$ or $s_1(\kappa)$, only one configuration exists, corresponding either to selecting the largest or the smallest eigenvalues. Therefore, $\P(s_{0}(\kappa)) \simeq 1/\binom{N}{\Nt}$. In the large $N$ limit,
\begin{equation}
	-\frac{1}{N} \ln \P(s_{0}(\kappa)) \simeq \frac{1}{N} \ln \binom{N}{\Nt}
		\simeq-\kappa \ln \kappa - (1-\kappa) \ln (1-\kappa)
		\equiv  C_\kappa 
	\:,
	\label{eq:447}
\end{equation}
and similarly for $s_1(\kappa)$.
The function $\Phi_0(\kappa;s)$ is plotted in Fig.~\ref{fig:LDFAsympt}.

%%%%%%%%%%%%%%%%%%%%%%%%%%%%%%%%%%%%%%%%%%%%%%%%%%%%%%%%%%%%%%%%%%%%%%%%%%%%%%%%%%%%%%%%%%
%%%%%%%%%%%%%%%%%%%%%%%%%%%%%%%%%%%%%%%%%%%%%%%%%%%%%%%%%%%%%%%%%%%%%%%%%%%%%%%%%%%%%%%%%%

\section{Phase III ($s>s_1(\kappa)$)~: mixed picture}
\label{sec:LDFlarge}

The distribution of the truncated linear statistics (\ref{eq:defTrlinstat0}) for $s < s_1(\kappa)$ was determined by considering two different situations.
First, for $s < s_0(\kappa)$, the fraction $\kappa$ of the largest eigenvalues detach from the others, as we have seen in section \ref{sec:Coulombgas}. When $s$ is increased, these two bulks of eigenvalues move closer, until they merge for $s = s_0(\kappa)$ (Fig.~\ref{fig:MPminmax}, left). Then, when $s$ is further increased in the interval $[s_0(\kappa),s_1(\kappa)]$, the density of eigenvalues is frozen and the occupation numbers $\lbrace \m_i \rbrace$ fluctuate, as it was shown in section \ref{sec:TypFluct}. When $s$ reaches $s_1(\kappa)$, these occupation numbers no longer fluctuate and only the smallest eigenvalues are selected (Fig.~\ref{fig:MPminmax}, right).

In the domain $s > s_1(\kappa)$ we could naively expect a similar scenario as for the tail $s \to 0$, namely to detach the fraction $\kappa$ of the smallest eigenvalues.
However, in the case $f(x) = 1/x$, there is another scenario more favorable energetically.
It is enough to detach one eigenvalue (the smallest, $x_N$) to obtain large values of $s$~:
a ``bulk'' of $N-1$ eigenvalues can remain frozen while only one eigenvalue contributes to the $s$-dependent part of the energy, which thus scales as $\sim N$ instead of $\sim N^2$ as in Phase~I.
We write
\begin{equation}
	s = \frac{1}{N x_N} + \frac{1}{N} \sum_{i=1}^{N-1} \frac{\m_i}{x_i}
	\:.
	\label{eq:sSplitCharge}
\end{equation}
If the smallest eigenvalue $x_N$ scales as $N^{-1}$, it gives a ``macroscopic'' contribution of the same order as the other $N-1$ eigenvalues which 
remain of order $N^0$. 
Because $x_N$ is much smaller that the other eigenvalues in the limit $N \to \infty$,
the constraint $x_N < x_{N-1}$ in the multiple integrals \eqref{eq:PnsRatioMultInt} (for $f(\lambda)=1/\lambda$) plays no role.
We estimate the integral over $\lbrace x_1, \cdots, x_{N-1} \rbrace$ with a saddle point method. The saddle point denoted $\lbrace x_i^\star \rbrace$, corresponds to the minimum of the energy $ \mathcal{E}_\mathrm{gas}[\lbrace x_i \rbrace]$, Eq.~(\ref{eq:MinEdiscr}). Thus, these eigenvalues are frozen and distributed according to the Mar\v{c}enko-Pastur density $\rhoMP$. The corresponding energy is~:
\begin{equation}
	\mathcal{E}_\mathrm{gas}[\lbrace x_1^\star, \cdots x_{N-1}^\star, x_N \rbrace] \simeq 
		\mathscr{E}[\rhoMP]
		- \frac{1}{N} \ln x_N
		+ \frac{\mathscr{C}}{N}
	\:,
\end{equation}
where the constant $\mathscr{C} = -1 - 2 \ln 2$ arises from a careful treatment of $1/N$ corrections~\cite{GraMajTexUnp}.
%The same procedure can be carried out with the denominator, but this time with all the eigenvalues since they are not constrained.
Taking into account the contribution of the denominator in \eqref{eq:PnsRatioMultInt},
we obtain~:
\begin{equation}
	\P(s) \sim 
		\sum_{ \lbrace \m_i \rbrace }
		 \int \dd x_N \,
			\e^{\frac{\beta N}{2} (\ln x_N - \mathscr{C}) }
		\delta \left(
			s - \frac{1}{N x_N} - \frac{1}{N} \sum_{i=1}^{N-1} \frac{\m_i}{x_i^\star}
		\right)
		\,
		\ConstN
  		\:.
\end{equation}
%Note that we did not discuss the entropic term $\mathscr{S}$ of order $N^1$ which corrects the energetic term $\mathscr{E}[\rhoMP]$ because it cancels out between the numerator and the denominator.
Note that the substitution \eqref{eq:MeasurePathInt} in the multiple integrals in \eqref{eq:PnsRatioMultInt} give contributions of the entropy of the Coulomb gas $(1-\beta/2)\mathscr{S}[\rho]$ also of order $N^1$, however the leading term of such contributions are the same in the numerator and the denominator and thus cancel.
Integration over the last eigenvalue is straightforward~:
\begin{equation}
	\P(s) \sim 
		\sum_{\lbrace \m_i \rbrace } \exp \left\lbrace
			-\frac{\beta N}{2}
				\left[
				\ln \left(
				N s - \sum_{i=1}^{N-1} \frac{\m_i}{x_i^\star}
			\right)
			+ \mathscr{C} \right]
		\right\rbrace
		\ConstN
	\:.
\end{equation}
In order to compute this sum, let us rewrite it as
\begin{equation}
	\P(s) \sim
		\int \dd u \:
			\e^{-\frac{\beta N}{2} (\ln [N(s - u)] + \mathscr{C} )}
		\sum_{\lbrace \m_i \rbrace }
			\delta \left(
				u - \frac{1}{N} \sum_{i=1}^{N-1} \frac{\m_i}{x_i^\star}
			\right)
		\ConstN
	\:.
	\label{eq:PnsLargeSum}
\end{equation}
We now recognize the distribution of $s$ computed in section \ref{sec:TypFluct}~:
\begin{equation}
	\sum_{\lbrace \m_i \rbrace }
			\delta \left(
				u - \frac{1}{N} \sum_{i=1}^{N-1} \frac{\m_i}{x_i^\star}
			\right)
		\ConstN
		\sim e^{- N \Phi_0(\kappa;u)}
	\:,
\end{equation}
where $\Phi_0$ is given by (\ref{eq:LdfPhi0}). Therefore, (\ref{eq:PnsLargeSum}) reduces to
\begin{equation}
	\P(s) \sim \int \dd u \:
		\exp \left\lbrace -\frac{\beta N}{2} ( \ln [N(s - u)] + \mathscr{C} ) - N \Phi_0(\kappa;u)
			\right\rbrace
	\:.		
\end{equation}
This last integral can be computed using a saddle point method. The saddle point $\tilde{s}(s)$ is given by~:
\begin{equation}
	\frac{\beta}{2} \frac{1}{\tilde{s} - s}
		+ \left. \deriv{\Phi_0(\kappa;u)}{u} \right|_{\tilde{s}(s)}
		= 0
	\:.
	\label{eq:SaddlePtSepCharge}
\end{equation}
The quantity $\tilde{s}$ represents the contribution of the frozen bulk to the TLS. Its limiting behaviours are found by using \eqref{eq:Phi0closeToKappa} and \eqref{eq:Phi0closeToS1}, where the latter is rewritten $\Phi_0(\kappa;s)\simeq C_\kappa-\tilde{c}_+\sqrt{s_1(\kappa)-s}$.
Some elementary algebra gives
\begin{equation}
  \label{eq:LimitsStilde}
  \tilde{s}(s) \simeq
  \begin{cases}
    s_1(\kappa) - \left(\frac{\tilde{c}_+}{\beta}\right)^2 (s-s_1(\kappa))^2 & \mbox{for } s\to s_1(\kappa)^+ 
    \\[0.25cm]
    \kappa + \frac{\beta\kappa(1-\kappa)}{2(s-\kappa)} + \mathcal{O}(s^{-3})   
      & \mbox{for } s\to\infty    
  \end{cases} 
\end{equation}

Finally, the probability is given by~:
\begin{equation}
   \boxed{
	\P(s) 
	  \underset{N\to\infty}{\sim}
	N^{-\beta N/2}
	\exp \left\lbrace
		- N \left[\frac{\beta}{2} \Phi_+(\kappa;s) + \Phi_0(\kappa,\tilde{s}(s)) \right]
	\right\rbrace
	\hspace{0.25cm}
	\text{for}
	\hspace{0.25cm}
	 s > s_1(\kappa)
    }
    %    \:,
\end{equation}
where the large deviation function is 
\begin{equation}
	\Phi_+(\kappa;s) = N( \mathcal{E}_\mathrm{gas}[\lbrace x_n \rbrace] - \mathcal{E}[\rhoMP(x)])
			-\ln N
			= \ln (s - \tilde{s}(s))  +\mathscr{C}
		\:,
\end{equation}
where $\mathscr{C}=- 1 - 2 \ln 2$.
There are two different contributions to the distribution of $s$:
\begin{itemize}
	\item the energy of the isolated eigenvalue $x_N$, encoded in $\Phi_+$;
	\item the entropy associated to the possible configurations $\lbrace \m_i \rbrace$ associated to the $N-1$ frozen eigenvalues, encoded in $\Phi_0$.
\end{itemize}
The latter was not present in the case $\kappa=1$ studied in Ref.~\cite{TexMaj13} because all the occupation numbers were fixed to $\m_i = 1$ and do not bring any additional entropy.

\subsection{Behaviour near the edge $s_1(\kappa)$}

Using \eqref{eq:LimitsStilde}, $\tilde{s} \simeq s_1(\kappa)$, gives
\begin{equation}
	\Phi_+(\kappa;s) \simeq \ln(s - s_1(\kappa) )+ \mathrm{cste}
	\:,
\end{equation}
and
\begin{equation}
	\Phi_0(\kappa; \tilde{s}(s)) \to  C_\kappa 
    %-\kappa \ln \kappa - (1-\kappa) \ln (1-\kappa) 
	\:,
\end{equation}
where $C_\kappa>0$ was defined in Eq.~\eqref{eq:447}.
Therefore, the leading order term
\begin{equation} 
	\Phi_+(\kappa;s) + \frac{2}{\beta} \Phi_0(\kappa;\tilde{s}(s))
	\simeq
	 \ln (s - s_1(\kappa))
	 + \mathrm{cste}
	\quad
	\text{for}
	\quad
	s \to s_1(\kappa)
\end{equation}
is independent of the Dyson index $\beta$, which only appears in the constant.

\subsection{Tail $s \to \infty$}

The limiting behaviour \eqref{eq:LimitsStilde}, $\tilde{s} \simeq \kappa$, shows that the entropy no longer plays a role and the contribution of the energy reads~:
\begin{equation}
	\Phi_+(\kappa;s) \simeq \ln(s - \kappa)  + \mathcal{O}(s^{-2})
	\:,
	\hspace{0.5cm}
	\text{for } s \to \infty 
	\:,
\end{equation}
which is again independent of $\beta$.
This gives the right tail of the distribution~:
\begin{equation}
	\P(s) \underset{s \to \infty}{\sim} s^{-\beta N/2} 
	\:.
\end{equation}
We have recovered the tail obtained in Ref.~\cite{TexMaj13} for $\kappa=1$~: $P_{N,1}(s) \sim (s-1)^{-\beta N/2}$ for $s-1 \gg \sqrt{\ln N/N}$, as expected since this limit is described within the same scenario (one isolated charge and a frozen bulk).
From the physical point of view, this power law tail is interpreted as the manifestation of a very narrow resonance which dominates the sum of proper times.

%%%%%%%%%%%%%%%%%%%%%%%%%%%%%%%%%%%%%%%%%%%%%%%%%%%%%%%%%%%%%%%%%%%%%%%%%%%%%%%%%%%%%%%%%%
%%%%%%%%%%%%%%%%%%%%%%%%%%%%%%%%%%%%%%%%%%%%%%%%%%%%%%%%%%%%%%%%%%%%%%%%%%%%%%%%%%%%%%%%%%

\section{Conclusion}
\label{sec:Conclusion}

\subsection{A new universal scenario}

As briefly reviewed in the introduction, the study of linear statistics has played a very important role in random matrix theory. 
In this article we have introduced a new type of question by considering the statistical analysis of partial sums of eigenvalues $\sum_{i=1}^p f(\lambda_i)$, so-called ``\textit{truncated linear statistics}'' (TLS), with $\kappa=p/N<1$, where $N$ is the size of the matrices.
This question has been set within the Laguerre ensemble for $f(\lambda)=1/\lambda$.
The study of this particular model was motivated by the analysis of partial sums of proper time delays in chaotic scattering, which are characteristic times of the scattering process playing an important role in electronic transport.

The specificity of the problem has led us to introduce two sets of random variables, the ordered eigenvalues $\lambda_1>\lambda_2>\cdots>\lambda_N$ and the ``occupation numbers'' $\{\m_i\}_{i=1,\cdots,N}$, with $\m_i=0$ or $1$, leading to express the TLS as $s=\sum_{i=1}^N\m_i\,f(\lambda_i)$.
Correspondingly, we have introduced two different physical interpretations~:
\begin{itemize}
\item 
  the eigenvalues $\{\lambda_i\}$ can be interpreted as the positions of a one-dimen\-sio\-nal gas of particles with long range (logarithmic) interactions, as usual in such problems.
\item
  The variables $\{\m_i\}$ can be viewed as occupation numbers for $p$ fictitious non-interacting fermions in $N$ ``energy levels'' $\{ \varepsilon_i=f(\lambda_i) \}$.
\end{itemize}
We have determined the large $N$ behaviour of the distribution $\P (s)$ of the TLS,~\footnote{
  The numerical analysis of \S~\ref{subsec:numerics} has shown that the large $N$ result describes very well the distribution already for $N\gtrsim50$.
} which was shown to be controlled by an optimal configuration (solution of saddle point equations), characterised by the density $\rho^\star(x;\kappa,s)$ of eigenvalues (with $\lambda=N x$ for the Laguerre ensemble), while the optimal occupations are given by the Fermi-Dirac distribution for an effective temperature, Eq.~\eqref{eq:FermiDirac}.
We have shown that varying the two parameters $(\kappa,s)$ drives phase transitions in this optimal configuration.

\paragraph{Phase II.---}

In the absence of any constraint on the eigenvalue density, the (most probable) density is the Mar\v{c}enko-Pastur distribution \eqref{eq:MarcenkoPastur} (Fig.~\ref{fig:MPminmax}).
For a fixed $\kappa$, this density can be associated to a whole spectrum of values of the TLS $s\in[s_0(\kappa),s_1(\kappa)]$, corresponding to different choices for the occupation numbers $\{\m_i\}$.
The two extreme situations correspond to choosing the $p$ largest or the $p$ smallest eigenvalues~: cf. Fig.~\ref{fig:MPminmax}. This phase exists in a domain of the phase diagram bounded by two lines $s_1(\kappa)$ and $s_0(\kappa)$ (Fig.~\ref{fig:PhDiag}).
The energy of the Coulomb gas is frozen and the structure of the distribution is entirely due to the \textit{entropy} of the $p$ fictitious non-interacting fermions, which scales as $\sim N$ and is independent of the symmetry index $\beta$, thus $\P (s)\sim\exp \left\lbrace - N \Phi_0(\kappa;s) \right\rbrace$, where the large deviation function $\Phi_0(\kappa;s)$ has been determined in Section~\ref{sec:TypFluct}. 
As a consequence of this scaling, the relative fluctuations of the TLS are of order $\sim1/\sqrt{N}$ (whereas full linear statistics usually present relative fluctuations $\sim1/N$ as a consequence of the strong correlations in the Coulomb gas).
As the limits $s\to s_0$ and $s\to s_1$ correspond to select the contributions of the largest or the smallest eigenvalues (Fig.~\ref{fig:MPminmax}), they are associated to effective (absolute) fermionic temperature $1/\tilde{\beta}$ positive or negative, respectively (see also Fig.~\ref{fig:MPminmaxFD}).
This new scenario is completly universal (independent of the function $f$ and the specific matrix ensemble).

\paragraph{Phase I.---}

When $s$ reaches the lower boundary $s_0(\kappa)$ of the central domain of the phase diagram (Fig.~\ref{fig:PhDiag}), the effective temperature is fixed to $0^+$ and only the largest eigenvalues contribute to the TLS (Fig.~\ref{fig:MPminmax}).
Smaller value $s<s_0(\kappa)$ can only be realised by a deformation of the optimal density $\rho^\star(x;\kappa,s)$, which is split in two bulks, while occupations are frozen ($1/\tilde{\beta}=0^+$).
The weight of the optimal configuration is now controlled by the \textit{energy} of the Coulomb gas.
In the $s=\sum_{i=1}^p\lambda_i^{-1}\to0$ limit, we have $\lambda_i\sim p/s$, hence the energy is dominated by the confinment energy~: $E_N=N^2\mathcal{E}_\mathrm{gas}\sim p^2/s$.
This corresponds to the behaviour $\P (s)\sim\exp[-(\beta/2)N^2\Phi_-(\kappa;s)]$ with $\Phi_-(\kappa;s)\simeq\kappa^2/s$ (see Section~\ref{sec:Coulombgas}).

\paragraph{Phase III.---}

The case where $s>s_1(\kappa)$ was analysed with a mixed scenario.
As for $\kappa=1$ \cite{TexMaj13}, one remarks that it is sufficient to split off a single eigenvalue from the bulk, the smallest one $\lambda_N$, in order to generate large values of the TLS $s=1/\lambda_N+\sum_{i=1}^{N-1}n_i/\lambda_i$~:
i.e. it is more favorable energetically to send one charge towards $0$ as $s$ grows, which generates an energy cost $E_N\sim N$, than move the full bulk, which would generate a cost $E_N\sim N^2$.
When the charge is split off the bulk, there is also an entropic contribution $\sim N$ due to the fluctuating $N-1$ occupation numbers related to the remaining $N-1$ eigenvalues, of the same order as the interaction energy between the isolated eigenvalue and the frozen $N-1$ ones.
This part of the scenario is specific to the function $f(\lambda)=1/\lambda$ that we have considered. 

\vspace{0.25cm}

For the case considered in the paper, we have obtained three phases (Fig.~\ref{fig:PhDiag}), however one can imagine that for another function $f$, the phase diagram would involve more than three phases~: the exhaustive study  of Ref.~\cite{VivMajBoh10} for several linear statistics illustrates than the number of different phases depend on the function~$f$ (see also the Table in the conclusion of Ref.~\cite{GraMajTex17}).

% ------------------------------------------------------------------------------------------
\subsection{Crossover across $s_0(\kappa)$ and second order freezing transition}

As the two phases I and II involve two different scalings with $N$, see \eqref{eq:SummarisePN}, it is interesting to discuss further the transition at $s=s_0$, in the vicinity of which the distribution presents the limiting behaviours 
\begin{equation}
	\P(s) \underset{s \to s_0}{\sim}
	\e^{-N C_\kappa}
	\left\lbrace
		\begin{array}{lc}
		    \e^{-(\beta/2)N^2\Phi_-(\kappa;s)}
			\simeq
			\e^{- N^2 \cm (s - s_0(\kappa))^2}
				& \text{ for } s < s_0(\kappa) 
				\\
		    \e^{-N[\Phi_0(\kappa;s)-C_\kappa]}
			\simeq
			\e^{+N \cp \sqrt{s - s_0(\kappa)}}
				& \text{ for } s > s_0(\kappa)  
		\end{array}
	\right.
	\label{eq:PnkTr}
	\:,
\end{equation}
where $\cm=(\beta/2)\omega_\kappa$, cf. Eq.~\eqref{eq:LDFphi0QuadMin}, and $\cp$ is defined in Eq.~\eqref{eq:LimitingBehaviour344}.
From the point of view of the Coulomb gas, $s=s_0(\kappa)$ corresponds to a \textit{freezing transition}~: below $s_0(\kappa)$, the energy scales as $E_N\sim N^2 (s - s_0(\kappa))^2$, whereas the energy is frozen above the transition. This corresponds to a \textit{second order} phase transition (a second order freezing transition of different nature was identified in \cite{TexMaj13}, and also in \cite{NadMajVer10,NadMajVer11} although the transition was not interpreted along these lines in this latter reference).
Note that the energy of the Coulomb gas, Eq.~\eqref{eq:Ediscr}, is $\mathcal{E}_\mathrm{gas}\sim1/N$ everywhere above the line $s_0(\kappa)$. In the thermodynamic limit $N\to\infty$, $\mathcal{E}_\mathrm{gas}$ becomes flat above this line and all derivatives of the energy are trivially continuous across $s_1(\kappa)$.

A finer description of the transition (beyond the thermodynamic limit) requires to identify the non trivial scaling with $N$ for the crossover, given by 
\begin{equation}
	N^2 (s - s_0(\kappa))^2 \sim N \sqrt{s - s_0(\kappa)}
	\hspace{0.5cm}
	\Rightarrow
	\hspace{0.5cm}
	s - s_0(\kappa) \sim N^{-2/3}
	\:.
\end{equation}
This leads to introduce a scaling variable $z$ and rewrite the crossover between the two behaviours~\eqref{eq:PnkTr} in terms of a unique function
\begin{equation}
	\P(s) \underset{s \to s_0(\kappa)}{\sim} \e^{- N^{2/3} \Psi(z)} 
	\hspace{1cm}\mbox{with }
	s = s_0(\kappa) + N^{-2/3} z 
	\:.
\end{equation}
The scaling function $\Psi(z)$ describes the detail of the crossover for finite $N$, at a scale $\delta s\sim N^{-2/3}$, and thus presents the limiting behaviours
\begin{equation}
	\Psi(z) \simeq \left\lbrace
		\begin{array}{cc}
			\cm z^2	    &  \text{ for } z \to-\infty \:,
			\\
			-\cp \sqrt{z}	&  \text{ for } z \to+\infty \:.
		\end{array}
	\right.
\end{equation}
Similar considerations for the distribution of the largest eigenvalue in various matrix ensembles (see the review \cite{MajSch14}) have shown that the corresponding function is universal (Tracy Widom distribution given by Painlev\'e transcendents).
An interesting challenging open question would be to determine whether the function $\Psi(z)$ has also a universal character and find its precise nature.

\subsection{Other open questions}
\label{subsec:OpenQuestions}

The discussion concerning the crossover around $s_0(\kappa)$ cannot be extended to the other phase boundary at $s_1(\kappa)$, as it is not clear how the limiting behaviours obtained above,
$-(1/N)\ln\P(s)\simeq C_\kappa-\cp\sqrt{s_1-s}$ for $s<s_1$
and 
$-(1/N)\ln\P(s)\simeq C_\kappa+(\beta/2)\ln\big[N(s-s_1)\big]$ for $s>s_1$,
can be matched.
Due to the $N$ in the argument of the logarithm, it does not seem that the crossover can be described by a universal function as for $s\sim s_0$.
A precise description of the crossover seems therefore even more challenging in this case.

Another obvervation concerning finite $N$ corrections to the thermodynamics properties is the following~:
our approach is suitable to obtain results in the thermodynamic limit, when both $N$ and $p$ are of the same order (in particular the case $p=1$ would require to adapt the method like it was done in Refs.~ \cite{DeaMaj06,DeaMaj08,VivMajBoh07,MajVer09,MajSchVilViv13,MajSch14}). 
Also the limit $\kappa\to1$ is non trivial and can lead to singular behaviours, as it was illustrated in the conclusion of \cite{GraMajTex17}. This is clear from our fermionic picture as fermions and holes play symmetric role, so that the two limits $\kappa\sim1/N$ and $1-\kappa\sim1/N$ should be equally difficult to analyse. 

The new universal scenario introduced in the paper was illustrated for a particular ensemble (Laguerre) and a specific \textit{monotonic} function $f$.
We argue now that the same main features would also occur for a \textit{non monotonic} function.
Consider for concreteness the Jacobi case, $\lambda_i\in[0,1]$ and the function $f(\lambda)=\lambda\,(1-\lambda)$ (when $\sum_{i=1}^N\lambda_i(1-\lambda_i)$ is the shot noise power of chaotic quantum dots \cite{Bee97}).
In Phase II, due to the fact that the ``energy levels'' $\varepsilon_i=f(\lambda_i)$ are non monotonic as a function of the eigenvalues of the random matrix, the occupations controlled by the Fermi-Dirac distribution $\mean{\m_i}$ would also be represented by a non monotonic function of $\lambda$ (a similar feature occurs when representing the occupation of plane waves by one-dimensional non-interacting fermions~: occupation is a  monotonic function of the energy $\varepsilon_k=k^2$ but a non monotonic function of the wave vector $k$).
The limit $s=(1/N)\sum_i\m_if(\lambda_i)\to s_0^+$ (lower boundary of Phase II) should correspond to ``cold fermions'' with $\tilde{\beta}\to+\infty$ where the occupation function $\mean{\m_i}$ selects both the smallest and the largest eigenvalues. 
The large deviations for $s<s_0$ (Phase I) should correspond then to the opening of two gaps.
The upper boundary of Phase II, $s\to s_1^-$, corresponds to ``hot fermions'' with negative temperature (i.e. $\tilde{\beta}\to-\infty$) involving the eigenvalues in the middle of the interval $[0,1]$~; the large deviations for $s>s_1$ should also correspond to the opening of two gaps. 
Such a situation for a non monotonic $f$ could certainly be worth studying in details.

We have pointed out the connection between the problem studied in the paper and the ``thinned ensembles'' \cite{BohPat06,ChaCla16,BerDui16,Lam16}, Eq.~\eqref{eq:RelationWithThinned}.
It is clear that the correspondence between the two situations (fixed \textit{versus} fluctuating $p$) requires $\kappa=\Nt/N$ being the probability of the thinned ensemble.
Because the large deviation function controlling the distribution of the truncated linear statistics  has the interpretation of a thermodynamic potential and has been computed in the thermodynamic limit, one could naively expect that the two situations are equivalent.
However, the equivalence of the statistical physics ensembles is far from being obvious in these log correlated gases and may be worth to explore more into detail (note that, in our paper, the equivalence between grand canonical, canonical and microcanonical ensembles has been extensively used only for the \textit{non interacting} fictitious \textit{fermions}, i.e. only in the central region of the phase diagram (Fig.~\ref{fig:PhDiag}) when eigenvalues are \textit{frozen}). 

Coming back to the initial motivation within chaotic scattering, the paper has mostly focused on the question of the partial sums of proper time delays.
We have also emphasized the role of the Wigner-Smith matrix' diagonal elements $\WSm_{ii}$, which are important for applications in quantum transport~: 
the statistical properties of the partial sum of these matrix elements were analysed in the appendix for $\beta=2$, based on the relation between $\WSm_{ii}$ and partial time delays $\tilde{\tau}_i$ \cite{SavFyoSom01}. 
Because the joint distribution of the $\WSm_{ii}$'s is still unknown, a general analysis of their partial sums in the general case, beyond the two first moments, remains a challenging question.

\section*{Acknowledgements}

We are indebted to Dmitry Savin for many useful discussions on chaotic scattering.

%%%%%%%%%%%%%%%%%%%%%%%%%%%%%%%%%%%%%%%%%%%%%%%%%%%%%%%%%%%%%%%%%%%%%%%%%%%%%%%%%%%%%%%%%%
%%%%%%%%%%%%%%%%%%%%%%%%%%%%%%%%%%%%%%%%%%%%%%%%%%%%%%%%%%%%%%%%%%%%%%%%%%%%%%%%%%%%%%%%%%
%%%%%%%%%%%%%%%%%%%%%%%%%%%%%%%%%%%%%%%%%%%%%%%%%%%%%%%%%%%%%%%%%%%%%%%%%%%%%%%%%%%%%%%%%%

\appendix

%%%%%%%%%%%%%%%%%%%%%%%%%%%%%%%%%%%%%%%%%%%%%%%%%%%%%%%%%%%%%%%%%%%%%%%%%%%%%%%%%%%%%%%%%%
%%%%%%%%%%%%%%%%%%%%%%%%%%%%%%%%%%%%%%%%%%%%%%%%%%%%%%%%%%%%%%%%%%%%%%%%%%%%%%%%%%%%%%%%%%

\section{Distribution of the injectance for $\beta=2$}
\label{app:Injectance}

In this section, we derive the distribution of the injectance of a quantum dot defined in section \ref{sec:QuantScatt}~:
\begin{equation}
	\overline{\nu}_\alpha= \frac{1}{2\pi} \sum_{i\in\mathrm{contact}\:\alpha}\WSm_{ii}
	\:.
\end{equation}
We will restrict our study to the situation where $\beta=2$ because the joint distribution of the $\WSm_{ii}$'s is known only in this case.
We will derive the distribution of $\overline{\nu}_\alpha$ using a Coulomb gas approach.
The computation is very similar to the one described in \cite{TexMaj13} where the Wigner time delay (case where all contacts contribute) is studied.
We start from the distribution of a subblock $\WSm_{\Nt}$ of size $ \Nt \times \Nt$ on the diagonal of $\WSm$ given in \cite{SavFyoSom01} for $\beta=2$~:
\begin{equation}
	P(\WSm_{\Nt}^{-1}) \propto (\det \WSm_{\Nt})^{-N} \e^{- \tr{ \WSm_{\Nt}^{-1}}} \:.
\end{equation}
If we suppose that the contact $\alpha$ has $\Nt$ open channels, we can rewrite the injectance as~:
\begin{equation}
	\overline{\nu}_\alpha= \frac{1}{2\pi}  \tr{ \WSm_{\Nt}} \:.
\end{equation}
Denoting $\lbrace \lambda_i \rbrace$ the eigenvalues of $\WSm_{\Nt}^{-1}$, we can write
\begin{equation}
	\overline{\nu}_\alpha = \frac{1}{2\pi} \sum_{i=1}^{\Nt} \frac{1}{\lambda_i} \:,
\end{equation}
where the joint probability density function of the $\lbrace \lambda_i \rbrace$ is given by
\begin{equation}
	P(\lambda_1 ,\cdots , \lambda_{\Nt}) \propto
		\prod_{i<j} \abs{\lambda_i - \lambda_j}^2 \prod_{i=1}^{\Nt} \lambda_i^{N} \e^{-\lambda_i} \:.
\end{equation}
%We will study the distribution of $\overline{\nu}_\alpha$ using Coulomb gas method (details are given in section \ref{sec:Coulombgas}).
Rescale the eigenvalues as $\lambda_i = \Nt x_i$ and introduce the empirical density
\begin{equation}
	\rho(x) = \frac{1}{\Nt} \sum_{i=1}^{\Nt} \delta(x - x_i) \:.
\end{equation}
The distribution of
\begin{equation}
	s = 2\pi \overline{\nu}_\alpha = \frac{1}{\Nt} \sum_{i=1}^{\Nt} \frac{1}{x_i}
		= \int \frac{\rho(x)}{x} \dd x
\end{equation}
is given by
\begin{equation}
	\P(s) \simeq
	\frac{ \displaystyle
		\int \mathcal{D}\rho \: \e^{-\kappa^2 N^2 \mathscr{E}[\rho]} \:
		\delta \left(\int \rho(x)  \dd x - 1 \right)
		\delta \left(\int \frac{\rho(x)}{x} \dd x - s \right)
	}{ \displaystyle
		\int \mathcal{D}\rho \: \e^{- \kappa^2 N^2 \mathscr{E}[\rho]} \:
		\delta \left(\int \rho(x) \dd x - 1 \right)
	} \:,
	\label{eq:distInject}
\end{equation}
where we denoted $\kappa = \Nt/N$ and introduced the energy (see section \ref{sec:Coulombgas})
\begin{equation}
	\mathscr{E}[\rho] = - \int \dd x \int \dd y \: \rho(x) \rho(y) \ln \abs{x-y}
		+ \int \dd x \left(x - \frac{1}{\kappa} \ln x \right) \rho(x)
	\:.
\end{equation}
The integrals in (\ref{eq:distInject}) are dominated by the minimum of the energy under the constraints imposed by the $\delta$-functions. Therefore, we introduce the functional
\begin{equation}
  \label{eq:FCoulomb}
	\mathscr{F}[\rho,\mu_0,\mu_1] = \mathscr{E}[\rho]
		+ \mu_0 \left( \int \rho - 1 \right)
		+ \mu_1 \left( \int \frac{\rho(x)}{x} \dd x - s \right)
	\:,
\end{equation}
where $\mu_0$ and $\mu_1$ are Lagrange multipliers. Denote $\rho^\star(x;\kappa,s)$ the density which dominates the numerator of (\ref{eq:distInject}). It is given by $\left. \frac{\delta \mathscr{F}}{\delta \rho} \right|_{\rho^\star} = 0$, which reads~:
\begin{equation}
	2 \int \rho^\star(y;\kappa,s) \ln \abs{x-y} \dd y
		= x - \frac{1}{\kappa} \ln x + \mu_0 + \frac{\mu_1}{x}
		\quad \text{for} \quad x \in \mathrm{Supp}(\rho^\star)
	\:.
	\label{eq:MinFapp}
\end{equation}
It is more convenient to derive this relation with respect to $x$~:
\begin{equation}
	2 \dashint \frac{\rho^\star(x;\kappa,s)}{x-y} \dd y = 
	1 - \frac{1}{\kappa\, x} - \frac{\mu_1}{x^2}
	\quad \text{for} \quad x \in \mathrm{Supp}(\rho^\star)
	 \:.
	 \label{eq:MinFappDer}
\end{equation}
The values $\mu_0^\star$ and $\mu_1^\star$ taken by the Lagrange multipliers are fixed by imposing the constraints
\begin{equation}
	\int \rho^\star(x;\kappa,s) \dd x = 1 \:,
	\hspace{1cm}
	\int \frac{\rho^\star(x;\kappa,s)}{x} \dd x = s \:.
	\label{eq:AppConst}
\end{equation}
Similarly, denote $\rho_0^\star(x)$ the density which dominates the denominator. The distribution of $s$ is then given by~:
\begin{equation}
	\P(s) \sim \exp \left\lbrace
		- \kappa^2 N^2 \left(
			\mathscr{E}[\rho^\star(x;\kappa,s)]-\mathscr{E}[\rho_0^\star(x)]
		\right)
	\right\rbrace
	\:.
\end{equation}
The thermodynamic identity (\ref{eq:ThermoIdCoulomb}) mentioned in the body of the paper allows a direct computation of the energy via the relation~:
\begin{equation}
	\deriv{\mathscr{E}[\rho^\star(x;\kappa,s)]}{s} = - \mu_1^\star(\kappa;s) \:.
	\label{eq:ThermoIdApp}
\end{equation}
Our aim is now to compute the density $\rho^\star$.
As in the body of the paper, depending on the values of the parameters $\kappa$ and $s$, we will find different types of densities $\rho^\star$, which we interpret as different ``phases''.

\subsection{Phase I~: Solution with one compact support}

Let us assume that the solution $\rho^\star$ has one compact support $[a,b]$. Then, Eq.~(\ref{eq:MinFappDer}) can be solved using Tricomi's theorem \cite{Tri57}.  We obtain~:
\begin{equation}
	\rho^\star(x;\kappa,s) = \frac{x+c}{2\pi x^2} \sqrt{(x-a)(b-x)}
	\:.
\end{equation}
It is convenient to parametrize this solution in terms of $u = \sqrt{a/b}$. Then,
imposing $\rho^\star(a;\kappa,s)= \rho^\star(b;\kappa,s) = 0$, along with the constraints (\ref{eq:AppConst}) yields~:
\begin{equation}
	v = \sqrt{a b} = \frac{2u}{\kappa} \frac{(2\kappa+1)u^2 -2u + (2\kappa+1)}{(1-u^2)^2}\:,
	\label{eq:AppV}
\end{equation}
\begin{equation}
	\mu_1^\star =- \frac{4u^2}{\kappa^2} 
	\frac{((2\kappa+1)u^2 -2u + (2\kappa+1))(u^2 -2(2\kappa+1)u +1)}{(1-u^2)^4}\:,
	\label{eq:AppMu}
\end{equation}
\begin{equation}
	c = \frac{\mu_1^\star}{v}\:,
	\label{eq:AppC}
\end{equation}
\begin{equation}
	s = \sigma_\kappa(u) = 
		(1-u)^2
		\frac{-u^4 + 4 (3\kappa+1)u^3 + 2(4\kappa-3)u^2 + 4(3\kappa+1)u-1}
			{16u^2((2\kappa+1)u^2 -2u + (2\kappa+1))}
	\:.
	\label{eq:AppS}
\end{equation}
Given $\kappa$ and $s$, the last equation allows to compute $u$, from which all the other parameters are deduced.
One can check that in the limit $\kappa\to1$, one recovers the equations given in \cite{TexMaj13}.

The optimal density $\rho_0^\star$ for the denominator can be deduced from $\rho^\star$ by releasing the constraint (setting $\mu_1^\star = 0$).

\subsubsection{Domain of validity}

This solution exists as long as $\rho^\star$ is positive, which corresponds to $x+c \geq 0$. This gives the condition
\begin{equation}
	\kappa \geq \frac{1 - 3 u + 3 u^2 - u^3}{2 u (3 + u^2)}
	\:,
	\label{eq:lim_comp_ph}
\end{equation}
which can be rewritten as $s < s_{I}(\kappa)$. 
This corresponds to the lower domain delimited by the upper solid line on figure \ref{fig:Appdiag} (left).

\subsubsection{Typical fluctuations}

The typical fluctuations are controlled by the minimum of the energy, which is given by $\mu_1^\star=0$. The density is then
\begin{equation}
	\rho_0^\star(x) = \frac{\sqrt{(x-a_0)(b_0-x)}}{2\pi x} \:,
\end{equation}
where
\begin{equation}
	a_0 = \frac{2 \kappa +1 - 2\sqrt{\kappa(\kappa+1)}}{\kappa} \:,
		\quad
	b_0 = \frac{2 \kappa +1 + 2\sqrt{\kappa(\kappa+1)}}{\kappa} \:.
\end{equation}
The corresponding value of $s$ is $\kappa$, and expanding Eqs.~(\ref{eq:AppMu},\ref{eq:AppS}) for $s$ close to $\kappa$ yields~:
\begin{equation}
	\mu_1^\star = - \frac{s-\kappa}{\kappa^3(\kappa+1)} + O((s-\kappa)^2) \:.
\end{equation}
The energy is obtained by simple integration, via Eq.~(\ref{eq:ThermoIdApp})~:
\begin{equation}
	\mathscr{E}[\rho^\star(x;\kappa,s)]-\mathscr{E}[\rho_0^\star(x)] 
	= \frac{(s-\kappa)^2}{2\kappa^3(\kappa+1)} + \O((s-\kappa)^3) \:.
\end{equation}
Thus, the distribution of $s$ near $\kappa$ is given by~:
\begin{equation}
	\P(s) \sim \exp \left\lbrace
			- N^2 \frac{(s-\kappa)^2}{2\kappa(\kappa+1)}
		\right\rbrace
	\:.
\end{equation}
We recover the leading term of the variance given in the introduction, Eq.~(\ref{eq:VarianceSumQii})~:
\begin{equation}
	\Var(s) \simeq \frac{\kappa(1+\kappa)}{N^2}\:.
\end{equation}

\subsubsection{Limiting behaviour $s \to 0$}

The limit $s \to 0$ corresponds to $\mu_1^\star \to +\infty$. Expanding Eqs.~(\ref{eq:AppV},\ref{eq:AppMu},\ref{eq:AppC},\ref{eq:AppS}) in this limit gives~:
\begin{equation}
	\mu_1^\star = \frac{1}{s^2} - \frac{\kappa+2}{2\kappa} \frac{1}{s} + \O(1) \:.
\end{equation}
Using again Eq.~(\ref{eq:ThermoIdApp}), we obtain~:
\begin{equation}
	\mathscr{E}[\rho^\star(x;\kappa,s)] = 
		\frac{1}{s} + \frac{\kappa+2}{2\kappa} \ln s + \O(1) \:,
\end{equation}
thus~:
\begin{equation}
	\P(s) \sim s^{-N^2\kappa(\kappa+2)/2 } e^{- \kappa^2 N^2/s}
	\hspace{0.5cm} \text{for} \hspace{0.5cm}
	 s \to 0 \:.
\end{equation}

\subsection{Phase II~: Solution with an isolated eigenvalue}

As for the Wigner time delay \cite{TexMaj13} ($\kappa=1$) and in Section \ref{sec:LDFlarge}, we look for a solution with an isolated eigenvalue~:
\begin{equation}
	\rho^\star(x;\kappa,s) = \frac{1}{\Nt}\, \delta(x-x_1) + \tilde{\rho}(x)\:,
\end{equation}
with now $\int \tilde{\rho} = 1-1/\Nt$. The minimization of $\mathscr{F}$ with respect to $\tilde{\rho}$ reads~:
\begin{equation}
	2 \dashint \frac{\tilde{\rho}(y)}{x-y} \dd y = 1 - \frac{1}{\kappa x} - \frac{\mu_1}{x^2}
		- \frac{2}{\Nt} \frac{1}{x-x_1} \:,
	\hspace{0.5cm} \text{for} \hspace{0.5cm} x \in \mathrm{Supp}(\tilde{\rho}) \:.
	\label{eq:minF_rhot}
\end{equation}
And minimization with respect to $x_1$ gives
\begin{equation}
	2 \int \frac{\tilde{\rho}(y)}{x_1 - y} = 1 - \frac{1}{\kappa x_1} - \frac{\mu_1}{x_1^2} \:,
	\hspace{0.5cm} \text{for} \hspace{0.5cm} x \in \mathrm{Supp}(\tilde{\rho}) \:.
	\label{eq:minF_x1}
\end{equation}
In addition, the constraint becomes~:
\begin{equation}
	s = \frac{1}{\Nt x_1} + \int \frac{\tilde{\rho}(x)}{x} \dd x \:.
\end{equation}
For $x_1$ to give a ``macroscopic'' contribution to $s$, we must have $x_1 = \O(N^{-1})$. Then equation \eqref{eq:minF_x1} imposes $\mu_1 = -x_1/\kappa + \O(\Nt^{-2})$. Finally, at leading order in $\Nt$, we get~:
\begin{equation}
	\tilde{\rho}(x) = \rho_0^\star(x) + \O(\Nt^{-1}) \:,
\end{equation}
\begin{equation}
	x_1 = \frac{1}{\Nt(s-\kappa)} + \O(\Nt^{-2}) \:,
\end{equation}
\begin{equation}
	\mu_1^\star = - \frac{1}{\kappa p} \frac{1}{s - \kappa} + \O(\Nt^{-2}) \:.
\end{equation}

% *********************************************************************************************

\subsubsection{Domain of validity}

\begin{figure}
\begin{center}
	\includegraphics[width=0.45\textwidth]{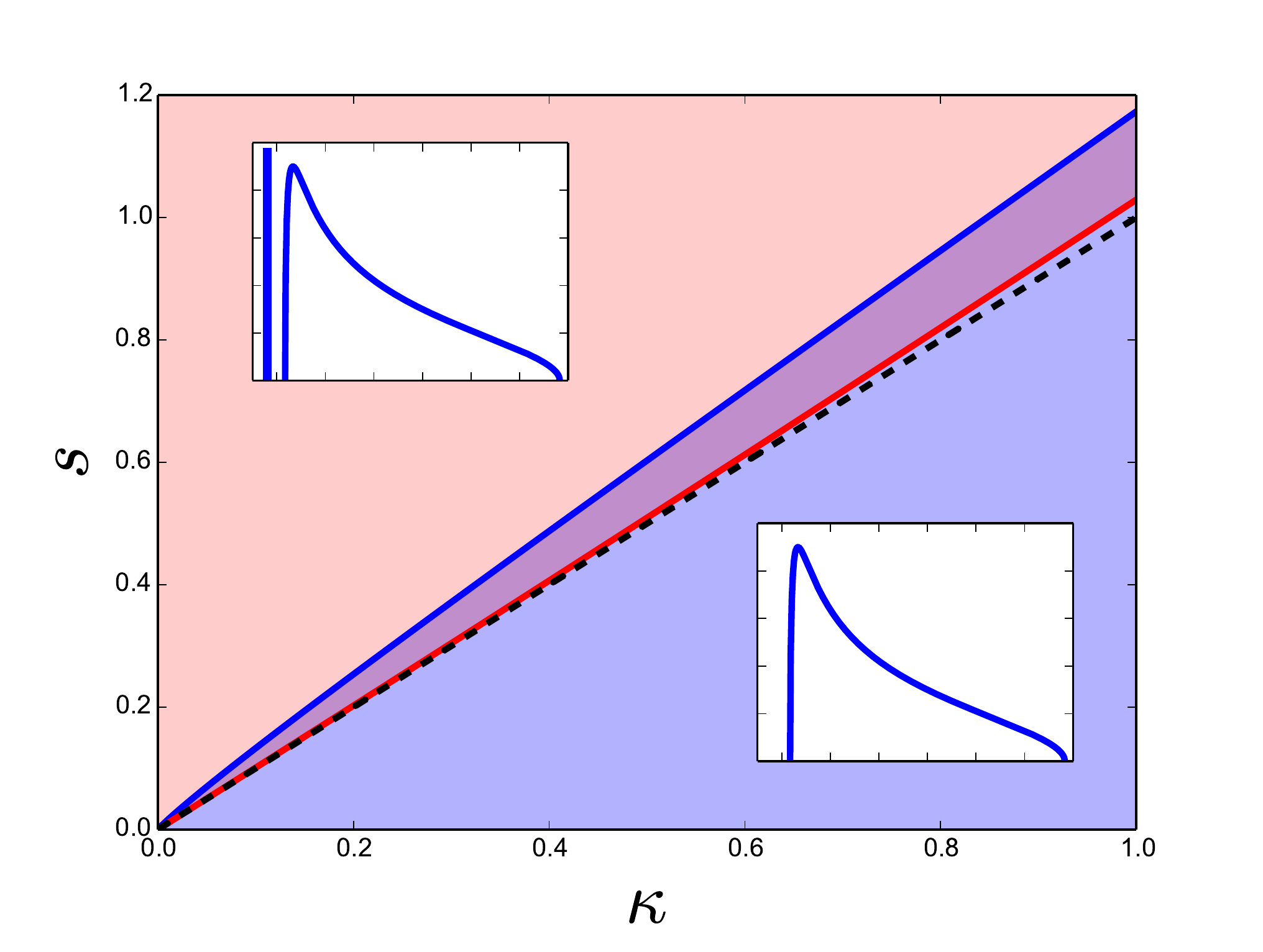}
	\includegraphics[width=0.45\textwidth]{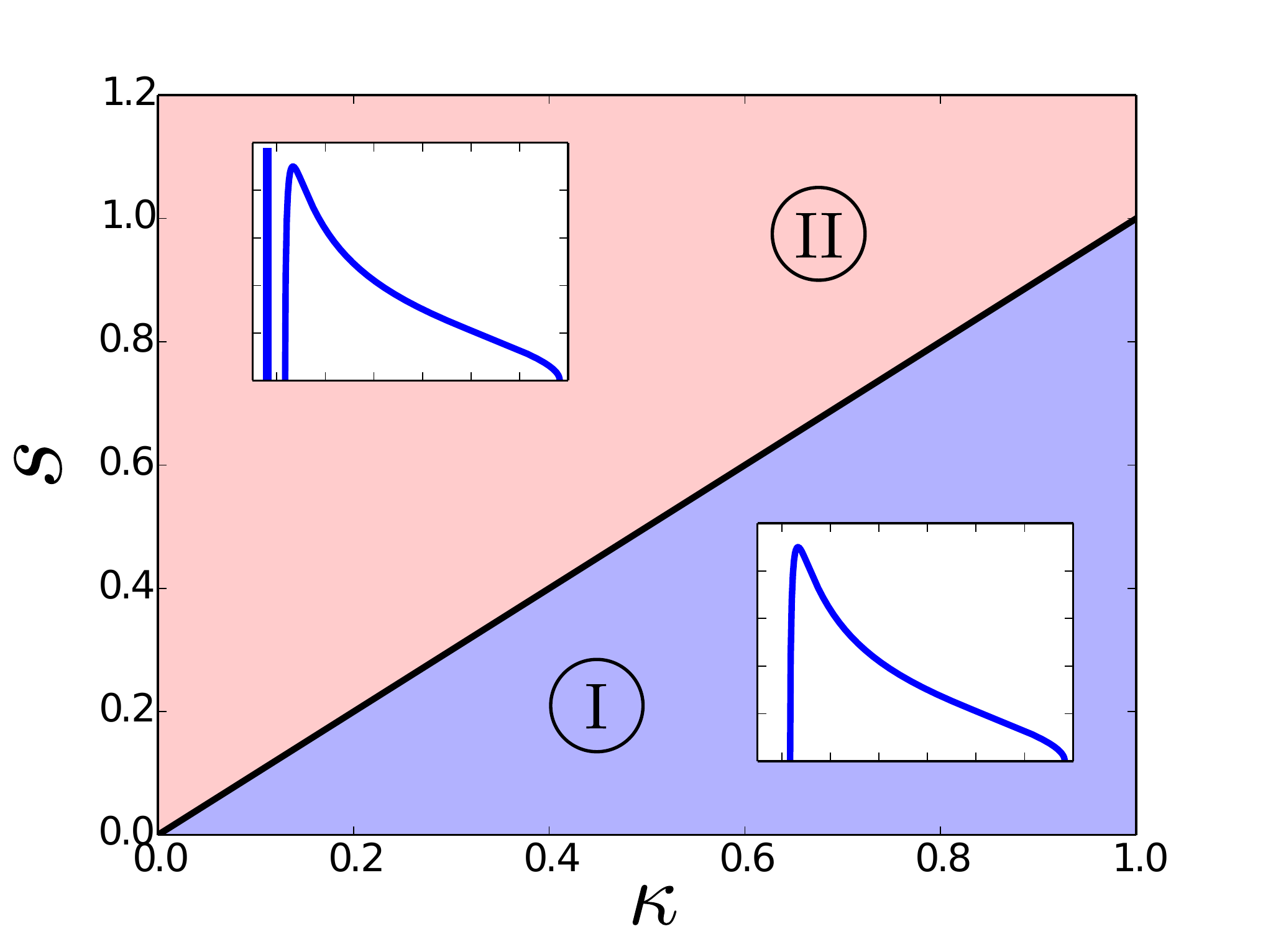}
\end{center}
\caption{Left: Domains of existence of each solution, for $N=200$.
The upper solid line corresponds to $s = s_{I}(\kappa)$ and delimits the domain of existence of phase I.
The lower solid line is $s = s_{II}(\kappa)$. Phase II exists above this line.
As $N \to \infty$, it goes to the dashed line $s = \kappa$,  where occurs the transition in the thermodynamic limit.
Right: phase diagram for the Coulomb gas.}
\label{fig:Appdiag}
\end{figure}

This solution remains valid as long as the separate eigenvalue is away from the bulk, namely $x_1 < a_0$. This gives the condition
\begin{equation}
	s > \kappa + \frac{1}{\Nt a_0} = s_{II}(\kappa) \:.
	\label{eq:lim_ph_ch}
\end{equation}
This corresponds to the upper domain represented on figure \ref{fig:Appdiag} (left).
Note that $s_{II}(\kappa) \to \kappa$ as $N \to \infty$.

% *********************************************************************************************

\subsubsection{Large deviation function}

The energy can be computed analytically at leading order~:
\begin{equation}
	\mathcal{E}[\rho^\star(x;\kappa,s)] - \mathcal{E}[\rho_0^\star]
	= \frac{1}{\kappa \Nt} \ln[\Nt (s - \kappa)]
		 + \frac{\mathscr{C}}{\Nt}
	+ \O(\Nt^{-2}).
	\label{eq:E_ph_sep_ch}
\end{equation}
where $\mathscr{C}=-1 - 2 \ln 2$ was introduced above.
The constant term is the same as the one appearing in section \ref{sec:LDFlarge} and Ref.~\cite{TexMaj13}. It arises from corrections of order $p^{-1}$ to the density $\rho^\star$ \cite{GraMajTexUnp}.
From the expression of the energy, we deduce the expression of the distribution of $s$, for $s > s_c$~:
\begin{equation}
	\P(s) \sim (s-\kappa)^{- N}
	\:.
\end{equation}
The simplification $\Nt^2/(\kappa \Nt)=N$ has thus produced the same exponent as for the tail of the distribution of the sum (truncated or not) of proper times.
This is explained from the interpretation of Ref.~\cite{SavFyoSom01}, where it was shown that $\WSm_{ii}$ coincides with the partial time delay for $\beta=2$.

% *********************************************************************************************

\subsection{Summary}

We obtained the following scalings for the distribution~:
\begin{equation}
	\P(s) \underset{N \to \infty}{\sim}
	\begin{cases}
    		\exp \left\lbrace - \kappa^2 N^2 \Psi_-(\kappa;s) \right\rbrace	
	\hspace{0.5cm}
   	 & \text{for }    s < s_{I}(\kappa) 
    \\[0.2cm]
    (\kappa N)^{-N}
   \exp \left\lbrace - N \Psi_+(\kappa;s) \right\rbrace	
	\hspace{0.5cm}
	& \text{for }    s > s_{II}(\kappa)
  \end{cases}
\end{equation}
where the large deviation $\Psi_+$ is given by~:
\begin{equation}
	\Psi_+(\kappa;s) = \ln(s - \kappa) - \kappa (1 + 2 \ln 2)\:,
\end{equation}
and $\Psi_-$ has the following limiting behaviours~:
\begin{equation}
	\Psi_-(\kappa;s) \simeq
	\begin{cases}
		\displaystyle
		\frac{1}{s} + \frac{\kappa+2}{2\kappa} \ln s + \O(1)
		\hspace{0.5cm}
		& \text{as } s \to 0
		 \\[0.2cm]
		 \displaystyle
		 \frac{(s-\kappa)^2}{2\kappa^3(\kappa+1)} + \O((s-\kappa)^3)
		 \hspace{0.5cm}
		& \text{as } s \to \kappa
	\end{cases}
\end{equation}

The precise point $s_t$ where the transition between the two phases occurs can be obtained by matching the two large deviation functions. One can show that, for $N \to \infty$, $s_t \to \kappa$.
Therefore for large $N$ we have, for $s$ close to $\kappa$~:
\begin{equation}
	\lim_{N \to \infty} - \frac{1}{N^2} \ln \P(s) \simeq
	\left\lbrace
		\begin{array}{cl}
			\displaystyle
			\frac{(s-\kappa)^2}{2\kappa^3(\kappa+1)}
			& 
			\text{ for } s < \kappa
			\\[0.3cm]
			0 & \text{ for } s > \kappa
		\end{array}
	\right.
\end{equation}
This corresponds to a \textit{second order} phase transition.
This was already the case in Ref.~\cite{TexMaj13} in the study of the full linear statistics ($\kappa=1$).

%%%%%%%%%%%%%%%%%%%%%%%%%%%%%%%%%%%%%%%%%%%%%%%%%%%%%%%%%%%%%%%%%%%%%%%%%%%%%%%%%%%%%%%%%%
%%%%%%%%%%%%%%%%%%%%%%%%%%%%%%%%%%%%%%%%%%%%%%%%%%%%%%%%%%%%%%%%%%%%%%%%%%%%%%%%%%%%%%%%%%

%\bibliographystyle{chris4}
%\bibliography{Biblio-Aurelien}
%\end{document}

\end{document}